# Photonic nanowire-based single-photon source with polarization control

## EXL02-REG2 Deliverable D1.3

**Niels Gregersen**

**19 May 2016**

**V1.0**

Abbreviations:

| | |
|---|---|
| EMRP | European Metrology Research Programme |
| IP | Intellectual Property |
| IPR | Intellectual Property Rights |
| IPX | Intellectual Property Exploitation |
| JRP | Joint Research Protocol |
| QKD | Quantum Key Distribution |
| REG | Research Excellence Grant |
| SIQUTE | Single-photon sources for quantum technologies |
| SPS | Single Photon Source |
| UK | United Kingdom |

# 1 Abstract

This document describes a modal method for optical simulations of structures with elliptical cross sections and its application to the design of the photonic nanowire (NW)-based single-photon source (SPS). The work was carried out in the framework of the EMRP SIQUTE project ending May 31st 2016. The document summarizes the new method used to treat the elliptical cross section in an efficient manner and additionally presents design parameters for the photonic NW SPS [1] with elliptical cross section for polarization control [2]. The document does not introduce the new method and the elliptical photonic NW SPS design in the context of existing literature but instead dives directly into the equations. Additionally, the document assumes that the reader possess expert knowledge of general modal expansion techniques [3]. The presented formalism does not implement Li's factorization rules [4] nor the recently proposed open boundary geometry formalism with fast convergence towards the open geometry limit [5] but instead relies on (older) formalism based on coordinate transformations [6] for emulating open systems. However, the document does contain all the information needed for the expert reader to implement the method and reproduce the simulations.

# 2 Introduction

From a numerical point of view, the photonic NW geometry is highly challenging due to the high index contrast between GaAs and due to the strong lateral scattering of light at the nanowire-metal boundary and the coupling to surface plasmons. While the FDTD method can be used for verification of the final design, the physics of the geometry in the design phase is generally analysed using a modal method.

To appreciate the challenge of the elliptical geometry and the strength of the new method, let us first consider the rotationally symmetric system. Here, the scalar optical field can be written as a sum

$$E_r(r,\phi,z) = \sum_m E_{r,m}(r,z)\cos(m\phi) \tag{1}$$

of decoupled contributions with different angular order *m*. Since the contributions are decoupled, it suffices to consider one order corresponding to the fundamental $HE_{11}$ waveguide mode. This effectively reduces the 3D calculation to a 2D simulation.

Now, the NW with an elliptical cross-section adds to the complexity of the calculation by breaking the rotational symmetry of the standard NW geometry and the problem can no longer be reduced to a 2D problem. Similarly to the case of the rotationally symmetric system, the optical field is expanded on natural basis functions, the Mathieu functions [7], of the elliptical geometry, known as the Fourier-Mathieu expansion basis. However, unlike the rotationally symmetric case, the Mathieu functions of different orders all couple to each other. Fortunately, only a limited number of orders are required to obtain convergence and this represents the advantage of the Fourier-Mathieu expansion method as compared to an expansion of standard plane waves in the Cartesian coordinate system.

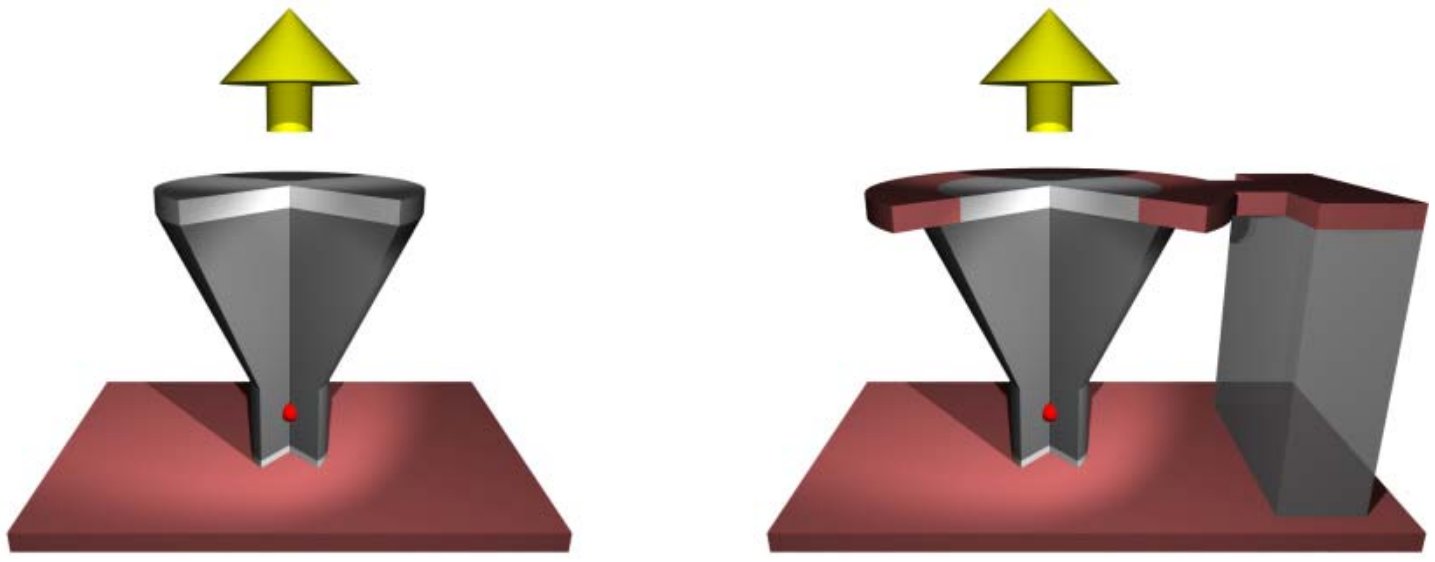

**Fig. 1: Artistic illustrations of the elliptical photonic nanowire geometries. Optically pumped design (left) and electrically pumped design (right) implementing a top metal contact.**

Optically and electrically pumped elliptical photonic NW geometries are illustrated in Fig. 1. Electrical contacting is implemented by adding contacts below and above the quantum emitter. Since the photonic nanowire design already features a bottom metal mirror, it is only necessary to add an annular gold ring contact at the taper top as illustrated in Fig. 1(right) [8].

After introducing the Fourier-Mathieu expansion method in Section 3, the method is used in the analysis of the photonic NW SPS with elliptical cross section for polarization control in Section 4.

## 3 Modal method using a Fourier-Mathieu expansion basis

For simplicity, we initially consider the scalar eigenvalue problem in the context of a generalized coordinate system. As specific example, we employ elliptical coordinates appropriate for the elliptical cross section of the NW. Subsequently, we present the full vectorial formalism for a generalized coordinate system.

### 3.1 Scalar eigenproblem

The starting point is the Helmholtz equation for the scalar electric field

$$\nabla^2 E(\mathbf{r}) + \varepsilon_r k_0{}^2 E(\mathbf{r}) = 0 . \tag{2}$$

In the eigenmode expansion technique, we have uniformity in the axis of propagation, that is $\varepsilon(\mathbf{r}_\perp)$ is independent of *z*. We thus use separation of variables

$$E(\mathbf{r}_\perp, z) = E(\mathbf{r}_\perp)\exp(i\beta z) , \tag{3}$$

leading to

$$\nabla_\perp^2 E(\mathbf{r}_\perp) + \varepsilon_r k_0{}^2 E(\mathbf{r}_\perp) = \beta^2 E(\mathbf{r}_\perp) . \tag{4}$$

We now introduce the general coordinate transformation $(x, y) \to (u, v)$, where the differential lengths are related as

$$l_u du = dx \tag{5}$$

$$l_v dv = dy , \tag{6}$$

where $l_u(u,v)$ and $l_v(u,v)$ are functions of *u* and *v*. The scalar wave equation becomes

$$\frac{1}{l_u l_v}\left(\frac{\partial}{\partial u}\frac{l_v}{l_u}\frac{\partial}{\partial u} + \frac{\partial}{\partial v}\frac{l_u}{l_v}\frac{\partial}{\partial v}\right)E(u,v) + \varepsilon_r k_0{}^2 E(u,v) = \beta^2 E(u,v) . \tag{7}$$

In the following we will often use the symbol Δ defined by

$$\Delta = \frac{1}{l_u l_v}\left(\frac{\partial}{\partial u}\frac{l_v}{l_u}\frac{\partial}{\partial u} + \frac{\partial}{\partial v}\frac{l_u}{l_v}\frac{\partial}{\partial v}\right) , \tag{8}$$

to represent the Laplace operator in generalized coordinates.

### 3.2 Elliptical coordinates

We now introduce the coordinate transformation for the elliptical coordinate system [7]

$$x = f\cosh(u)\cos(v) \tag{9}$$

$$y = f\sinh(u)\sin(v) , \tag{10}$$

with

$$0 \le u \le \infty \tag{11}$$

$$0 \le v \le 2\pi , \tag{12}$$

and

$$l_u = l_v = f\left[\frac{1}{2}\left(\cosh 2u - \cos 2v\right)\right]^{\frac{1}{2}} . \tag{13}$$

An ellipse is described by

$$\left(\frac{x}{r_x}\right)^2+\left(\frac{y}{r_y}\right)^2=1, \tag{14}$$

which corresponds to a surface of constant *u*, where *u* and *f* are related to $r_x$ and $r_y$ as

$$\cosh(2u)=\frac{r_x^2+r_y^2}{r_x^2-r_y^2} \tag{15}$$

$$f=\sqrt{r_x^2-r_y^2}\,. \tag{16}$$

The eigenvalue equation (4) in elliptical coordinates is

$$\frac{1}{f^2\left(\sinh^2 u+\sin^2 v\right)}\left(\frac{\partial^2}{\partial u^2}+\frac{\partial^2}{\partial v^2}\right)E(u,v)+\varepsilon(u)k_0^2E(u,v)=\beta^2E(u,v)\,. \tag{17}$$

We use separation of variables such that

$$E(u,v)=U(u)V(v)\,. \tag{18}$$

Insertion into Eq. (17) and division by *UV* gives

$$\frac{1}{\left(\sinh^2 u+\sin^2 v\right)}\left(\frac{1}{U}\frac{\partial^2 U}{\partial u^2}+\frac{1}{V}\frac{\partial^2 V}{\partial v^2}\right)=m^2\,, \tag{19}$$

where we have defined $m^2=f^2(\beta^2-\varepsilon k_0^2)$ where $\varepsilon$ is assumed to be a constant. Rearranging Eq. (19) gives

$$\left(\frac{1}{U}\frac{\partial^2 U}{\partial u^2}-m^2\sinh^2 u\right)+\left(\frac{1}{V}\frac{\partial^2 V}{\partial v^2}-m^2\sin^2 v\right)=0\,, \tag{20}$$

and separation of the two terms gives

$$\frac{\partial^2 U}{\partial u^2}-\left(c+m^2\sinh^2 u\right)U=0 \tag{21}$$

$$\frac{\partial^2 V}{\partial v^2}+\left(c-m^2\sin^2 v\right)V=0\,. \tag{22}$$

We now use

$$\sinh^2 u=\frac{1}{2}\left(\cosh(2u)-1\right) \tag{23}$$

$$\sin^2 v=\frac{1}{2}\left(1-\cos(2v)\right) \tag{24}$$

to obtain

$$\frac{\partial^2 U}{\partial u^2}-\left[\left(c-\frac{1}{2}m^2\right)+\frac{1}{2}m^2\cosh(2u)\right]U=0 \tag{25}$$

$$\frac{\partial^2 V}{\partial v^2}+\left[\left(c-\frac{1}{2}m^2\right)+\frac{1}{2}m^2\cos(2v)\right]V=0\,. \tag{26}$$

Using $a\equiv c-m^2/2$ and $q\equiv -m^2/4$, these equations become

$$\frac{\partial^2 V}{\partial v^2} + \left[a - 2q\cos(2v)\right]V = 0 \tag{27}$$

$$\frac{\partial^2 U}{\partial u^2} - \left[a - 2q\cosh(2u)\right]U = 0\,, \tag{28}$$

where Eq. (27) is the Mathieu differential equation defining the angular Mathieu function *V* and Eq. (28) is the modified Mathieu differential equation defining the radial Mathieu function *U*.

### 3.3 Evaluation of the field product integral for elliptical coordinates

The product integral is

$$\int E_1(x,y)E_2(x,y)dxdy = \int E_1(u,v)E_2(u,v)l_u l_v dudv\,. \tag{29}$$

In elliptical coordinates this becomes

$$\int \frac{f^2}{2}\left(\cosh(2u) - \cos(2v)\right)E_1(u,v)E_2(u,v)dudv\,. \tag{30}$$

We first rewrite Eq. (17) as

$$\frac{1}{f^2\left(\sinh^2 u + \sin^2 v\right)}\left(\frac{\partial^2}{\partial u^2} + \frac{\partial^2}{\partial v^2}\right)E(u,v) + k_r^2 E(u,v) = 0\,, \tag{31}$$

where we have introduced $k_r^2 = \varepsilon(u)k_0^2 - \beta^2$. Using Eqs. (23)-(24) and reorganizing we obtain

$$\left(\frac{\partial^2}{\partial u^2} + \frac{\partial^2}{\partial v^2}\right)E(u,v) + k_r^2 f^2 \frac{1}{2}\left(\cosh(2u) - \cos(2v)\right)E(u,v) = 0\,, \tag{32}$$

We have $q \equiv -m^2/4 = (fk_r)^2/4$ which leads to

$$\left(\frac{\partial^2}{\partial u^2} + \frac{\partial^2}{\partial v^2}\right)E(u,v) + 2q\left(\cosh(2u) - \cos(2v)\right)E(u,v) = 0\,. \tag{33}$$

We now assume two different solutions to Eq. (33) given by

$$\left(\frac{\partial^2}{\partial u^2} + \frac{\partial^2}{\partial v^2}\right)E_1(u,v) + 2q_1\left(\cosh(2u) - \cos(2v)\right)E_1(u,v) = 0 \tag{34}$$

$$\left(\frac{\partial^2}{\partial u^2} + \frac{\partial^2}{\partial v^2}\right)E_2(u,v) + 2q_2\left(\cosh(2u) - \cos(2v)\right)E_2(u,v) = 0\,. \tag{35}$$

We then form $E_2$·(34) - $E_1$·(35) leading to

$$E_2\left(\frac{\partial^2}{\partial u^2} + \frac{\partial^2}{\partial v^2}\right)E_1 - E_1\left(\frac{\partial^2}{\partial u^2} + \frac{\partial^2}{\partial v^2}\right)E_2 + 2(q_1 - q_2)\left(\cosh(2u) - \cos(2v)\right)E_1E_2 = 0\,, \tag{36}$$

which can be transformed into

$$\begin{aligned}&\frac{\partial}{\partial u}\left(E_2\frac{\partial}{\partial u}E_1 - E_1\frac{\partial}{\partial u}E_2\right) + \frac{\partial}{\partial v}\left(E_2\frac{\partial}{\partial v}E_1 - E_1\frac{\partial}{\partial v}E_2\right).\\ &+2(q_1 - q_2)\left(\cosh(2u) - \cos(2v)\right)E_1E_2 = 0\end{aligned} \tag{37}$$

Integration along *u* from $u_a$ to $u_b$ and along *v* from 0 to $2\pi$ gives

$$\int_0^{2\pi}\left[E_2\frac{\partial}{\partial u}E_1-E_1\frac{\partial}{\partial u}E_2\right]_{u_a}^{u_b}dv+\int_{u_a}^{u_b}\left[E_2\frac{\partial}{\partial v}E_1-E_1\frac{\partial}{\partial v}E_2\right]_0^{2\pi}du$$
$$+\int_{\substack{v=0\\u=u_a}}^{\substack{v=2\pi\\u=u_b}}2(q_1-q_2)\left(\cosh(2u)-\cos(2v)\right)E_1E_2dudv=0 \quad . \qquad (38)$$

The second integral cancels out due to periodicity leading to

$$\int_{\substack{v=0\\u=u_a}}^{\substack{v=2\pi\\u=u_b}}\left(\cosh(2u)-\cos(2v)\right)E_1E_2dudv=\frac{\int_0^{2\pi}\left[E_1\frac{\partial}{\partial u}E_2-E_2\frac{\partial}{\partial u}E_1\right]_{u_a}^{u_b}dv}{2(q_1-q_2)}. \qquad (39)$$

If $E_1 = U_1V_1$ and $E_2 = U_2V_2$, we can write

$$\int_{\substack{v=0\\u=u_a}}^{\substack{v=2\pi\\u=u_b}}\frac{f^2}{2}\left(\cosh(2u)-\cos(2v)\right)E_1E_2dudv=f^2\frac{\left[\left(U_1\frac{\partial}{\partial u}U_2-U_2\frac{\partial}{\partial u}U_1\right)\right]_{u_a}^{u_b}\int_0^{2\pi}V_1V_2dv}{4(q_1-q_2)}. \qquad (40)$$

If $E_1 = E_2 = UV$ we have to evaluate

$$\int_{\substack{v=0\\u=u_a}}^{\substack{v=2\pi\\u=u_b}}\frac{f^2}{2}\left(\cosh(2u)-\cos(2v)\right)U^2(u)V^2(v)dudv=$$
$$\int_{\substack{v=0\\u=u_a}}^{\substack{v=2\pi\\u=u_b}}\frac{f^2}{2}\left(\cosh(2u)U^2V^2-\cos(2v)U^2V^2\right)dudv \quad , \qquad (41)$$

using

$$\frac{1}{\pi}\int_0^{2\pi}ce_{2n}^2(v)\cos(2v)dv=A_0^{(2n)}A_2^{(2n)}+\sum_{r=0}^{\infty}A_{2r}^{(2n)}A_{2r+2}^{(2n)} \qquad (42)$$

$$\frac{1}{\pi}\int_0^{2\pi}ce_{2n+1}^2(v)\cos(2v)dv=\frac{1}{2}\left[A_1^{(2n+1)}\right]^2+\sum_{r=0}^{\infty}A_{2r+1}^{(2n+1)}A_{2r+3}^{(2n+1)} \qquad (43)$$

$$\frac{1}{\pi}\int_0^{2\pi}se_{2n+1}^2(v)\cos(2v)dv=-\frac{1}{2}\left[B_1^{(2n+1)}\right]^2+\sum_{r=0}^{\infty}B_{2r+1}^{(2n+1)}B_{2r+3}^{(2n+1)} \qquad (44)$$

$$\frac{1}{\pi}\int_0^{2\pi}se_{2n+2}^2(v)\cos(2v)dv=\sum_{r=0}^{\infty}B_{2r+2}^{(2n+2)}B_{2r+4}^{(2n+2)}, \qquad (45)$$

where $ce_n(v)$ and $se_n(v)$ are even (cosine-elliptic) and odd (sine-elliptic) angular Mathieu functions [9] respectively represented as expansions of corresponding cosine and sine functions with expansion coefficients $A_j^{(n)}$.

Furthermore, we have

$$\int_{u_a}^{u_b} U^2 \cosh(2u)du = \frac{1}{2}\left[\frac{\partial U}{\partial u}\frac{\partial U}{\partial q} - U\frac{\partial}{\partial u}\left(\frac{\partial U}{\partial q}\right)\right]_{u_a}^{u_b}. \tag{46}$$

However, we also need to evaluate

$$\int_{u=u_a}^{u=u_b} U^2 du\,, \tag{47}$$

and the procedure for analytically evaluating Eq. (47) is not clear. Let us instead write $E_2$ as a Taylor expansion in $\Delta q = q_2 - q_1$ as

$$E_2 = E_1 + \Delta q\frac{\partial E_1}{\partial q} + \ldots\,, \tag{48}$$

Negleting higher order terms in $\Delta q$, Eq. (39) then becomes

$$\begin{aligned} &\int_{\substack{v=0\\u=u_a}}^{\substack{v=2\pi\\u=u_b}} \left(\cosh(2u) - \cos(2v)\right) E_1\left(E_1 + \Delta q\frac{\partial E_1}{\partial q}\right) du dv \\ &= \frac{\int_0^{2\pi}\left[\left(E_1 + \Delta q\frac{\partial E_1}{\partial q}\right)\frac{\partial}{\partial u}E_1 - E_1\frac{\partial}{\partial u}\left(E_1 + \Delta q\frac{\partial E_1}{\partial q}\right)\right]_{u_a}^{u_b} dv}{2\Delta q} \end{aligned}. \tag{49}$$

Letting $\Delta q \to 0$ and skipping the $_1$ index we obtain

$$\int_{\substack{v=0\\u=u_a}}^{\substack{v=2\pi\\u=u_b}} \frac{f^2}{2}\left(\cosh(2u) - \cos(2v)\right) E^2 du dv = f^2\frac{\int_0^{2\pi}\left[\frac{\partial E}{\partial q}\frac{\partial}{\partial u}E - E\frac{\partial}{\partial u}\frac{\partial E}{\partial q}\right]_{u_a}^{u_b} dv}{4}. \tag{50}$$

Writing $E = UV$ this becomes

$$\begin{aligned} &\int_{\substack{v=0\\u=u_a}}^{\substack{v=2\pi\\u=u_b}} \frac{f^2}{2}\left(\cosh(2u) - \cos(2v)\right) E^2 du dv = f^2\frac{\int_0^{2\pi}\left[\left(\frac{\partial UV}{\partial q}\right)\frac{\partial}{\partial u}U - U\frac{\partial}{\partial u}\frac{\partial UV}{\partial q}\right]_{u_a}^{u_b} V dv}{4} \\ &= f^2\frac{\int_0^{2\pi}\left[\left(V\frac{\partial U}{\partial q} + U\frac{\partial V}{\partial q}\right)\frac{\partial}{\partial u}U - U\frac{\partial}{\partial u}\left(V\frac{\partial U}{\partial q} + U\frac{\partial V}{\partial q}\right)\right]_{u_a}^{u_b} V dv}{4} \\ &= f^2\frac{\int_0^{2\pi}\left[V\frac{\partial U}{\partial q}\frac{\partial U}{\partial u} + U\frac{\partial V}{\partial q}\frac{\partial U}{\partial u} - U\frac{\partial}{\partial u}\left(V\frac{\partial U}{\partial q}\right) - U\frac{\partial}{\partial u}\left(U\frac{\partial V}{\partial q}\right)\right]_{u_a}^{u_b} V dv}{4} \\ &= f^2\frac{\int_0^{2\pi}\left[\frac{\partial U}{\partial q}\frac{\partial U}{\partial u} - U\frac{\partial}{\partial u}\frac{\partial U}{\partial q}\right]_{u_a}^{u_b} V^2 dv}{4} \end{aligned}. \tag{51}$$

For different solutions $E_1 = U_1V_1$ and $E_2 = U_2V_2$ having different $a_1 \neq a_2$ but identical $q_1 = q_2 = q$, Eq. (27) becomes

$$\frac{\partial^2 V_1}{\partial v^2} + \left[a_1 - 2q\cos(2v)\right]V_1 = 0 \tag{52}$$

$$\frac{\partial^2 V_2}{\partial v^2} + \left[a_2 - 2q\cos(2v)\right]V_2 = 0 . \tag{53}$$

Forming $V_2\cdot(52)$ - $V_1\cdot(53)$ leads to

$$V_2\frac{\partial^2 V_1}{\partial v^2} - V_1\frac{\partial^2 V_2}{\partial v^2} + \left(a_1 - a_2\right)V_1V_2 = 0 , \tag{54}$$

which can be written as

$$\frac{\partial}{\partial v}\left(V_2\frac{\partial V_1}{\partial v} - V_1\frac{\partial V_2}{\partial v}\right) + \left(a_1 - a_2\right)V_1V_2 = 0 . \tag{55}$$

Integration from 0 to $2\pi$ then gives

$$\left[V_2\frac{\partial V_1}{\partial v} - V_1\frac{\partial V_2}{\partial v}\right]_0^{2\pi} + \int_0^{2\pi}\left(a_1 - a_2\right)V_1V_2 dv = 0 , \tag{56}$$

where the first term cancels out due to periodicity leaving

$$\int_0^{2\pi} V_1V_2 dv = 0 . \tag{57}$$

### 3.4 Vectorial eigenproblem

The curl and divergence operators in elliptical coordinates can be written as

$$\nabla\times\mathbf{A} = \left(\frac{1}{l_v}\frac{\partial A_z}{\partial v} - \frac{\partial A_v}{\partial z}\right)\mathbf{u} + \left(\frac{\partial A_u}{\partial z} - \frac{1}{l_u}\frac{\partial A_z}{\partial u}\right)\mathbf{v} + \frac{1}{l_u l_v}\left(\frac{\partial(l_v A_v)}{\partial u} - \frac{\partial(l_u A_u)}{\partial v}\right)\mathbf{z} \tag{58}$$

$$\nabla\cdot\mathbf{A} = \frac{1}{l_u l_v}\left(\frac{\partial(l_v A_u)}{\partial u} + \frac{\partial(l_u A_v)}{\partial v}\right) + \frac{\partial A_z}{\partial z} . \tag{59}$$

For a fixed frequency $\omega_0$ the Maxwell equations can be written as

$$\nabla\times\mathbf{E} = i\omega_0\mu_0\mathbf{H} \tag{60}$$

$$\nabla\times\mathbf{H} = -i\omega_0\varepsilon(r)\mathbf{E} . \tag{61}$$

We write the fields in the following form:

$$\mathbf{E}(u,v,z) = \mathbf{e}(u,v)\exp(i\beta z) \tag{62}$$

$$\mathbf{H}(u,v,z) = \mathbf{h}(u,v)\exp(i\beta z) \tag{63}$$

where $l \equiv l_u = l_v = f\left[\frac{1}{2}\left(\cosh 2u - \cos 2v\right)\right]^{\frac{1}{2}}$.

Maxwell's equations in elliptical coordinates are

$$i\omega_0\mu_0 h_u = \frac{1}{l_v}\frac{\partial e_z}{\partial v} - i\beta e_v \tag{64}$$

$$i\omega_0\mu_0 h_v = i\beta e_u - \frac{1}{l_u}\frac{\partial e_z}{\partial u} \tag{65}$$

$$i\omega_0\mu_0 h_z = \frac{1}{l_u l_v}\left(\frac{\partial(l_v e_v)}{\partial u} - \frac{\partial(l_u e_u)}{\partial v}\right) \tag{66}$$

$$-i\omega_0\varepsilon_0 e_u = \frac{1}{l_v}\frac{\partial h_z}{\partial v} - i\beta h_v \tag{67}$$

$$-i\omega_0\varepsilon_0 e_v = i\beta h_u - \frac{1}{l_u}\frac{\partial h_z}{\partial u} \tag{68}$$

$$-i\omega_0\varepsilon_0 e_z = \frac{1}{l_u l_v}\left(\frac{\partial(l_v h_v)}{\partial u} - \frac{\partial(l_u h_u)}{\partial v}\right). \tag{69}$$

Combining the equations gives

$$e_u = i\frac{\beta}{k_r^2 l_u}\frac{\partial e_z}{\partial u} + i\omega_0\mu_0\frac{1}{k_r^2 l_v}\frac{\partial h_z}{\partial v} \tag{70}$$

$$e_v = i\frac{\beta}{k_r^2 l_v}\frac{\partial e_z}{\partial v} - i\omega_0\mu_0\frac{1}{k_r^2 l_u}\frac{\partial h_z}{\partial u} \tag{71}$$

$$h_u = -i\omega_0\varepsilon_0\frac{1}{k_r^2 l_v}\frac{\partial e_z}{\partial v} + i\frac{\beta}{k_r^2 l_u}\frac{\partial h_z}{\partial u} \tag{72}$$

$$h_v = i\omega_0\varepsilon_0\frac{1}{k_r^2 l_u}\frac{\partial e_z}{\partial u} + i\frac{\beta}{k_r^2 l_v}\frac{\partial h_z}{\partial v}. \tag{73}$$

In the following we will generally use the scaling $\mathbf{H}\omega_0\mu_0 = \mathbf{H}'$, where we generally suppress the ' for simplicity. The empty geometry solutions are then given from the relations

$$e_u = \frac{1}{k_r l_u}\frac{\partial}{\partial u}U_q(u)V_q(v) \quad e_v = \frac{1}{k_r l_v}\frac{\partial}{\partial v}U_q(u)V_q(v) \quad e_z = k_r\frac{U_q(u)V_q(v)}{i\beta} \ \textbf{(TM)} \tag{74}$$

$$h_u = -\frac{\varepsilon_r k_0^2}{k_r\beta}\frac{1}{l_v}\frac{\partial}{\partial v}U_q(u)V_q(v) \quad h_v = \frac{\varepsilon_r k_0^2}{k_r\beta}\frac{1}{l_u}\frac{\partial}{\partial u}U_q(u)V_q(v) \quad h_z = 0 \ \textbf{(TM)} \tag{75}$$

$$e_u = \frac{k_0}{k_r\beta}\frac{1}{l_v}\frac{\partial}{\partial v}U_q(u)V_q(v) \quad e_v = -\frac{k_0}{k_r\beta}\frac{1}{l_u}\frac{\partial}{\partial u}U_q(u)V_q(v) \quad e_z = 0 \ \textbf{(TE)} \tag{76}$$

$$h_u = k_0\frac{1}{k_r l_u}\frac{\partial}{\partial u}U_q(u)V_q(v) \quad h_v = k_0\frac{1}{k_r l_v}\frac{\partial}{\partial v}U_q(u)V_q(v) \quad h_z = k_0 k_r\frac{U_q(u)V_q(v)}{i\beta} \ \textbf{(TE)} \tag{77}$$

We will use the basis functions given by

$$e_u = \frac{1}{k_r l_u}\frac{\partial}{\partial u}U_q(u)V_q(v) \quad e_v = \frac{1}{k_r l_v}\frac{\partial}{\partial v}U_q(u)V_q(v) \quad e_z = U_q(u)V_q(v) \ \textbf{(TM)} \tag{78}$$

$$h_u = -\frac{1}{k_r l_v}\frac{\partial}{\partial v}U_q(u)V_q(v) \quad h_v = \frac{1}{k_r l_u}\frac{\partial}{\partial u}U_q(u)V_q(v) \quad h_z = 0 \ \textbf{(TM)} \tag{79}$$

$$e_u = \frac{1}{k_r l_v}\frac{\partial}{\partial v}U_q(u)V_q(v) \quad e_v = -\frac{1}{k_r l_u}\frac{\partial}{\partial u}U_q(u)V_q(v) \quad e_z = 0 \ \textbf{(TE)} \tag{80}$$

$$h_u = \frac{1}{k_r l_u}\frac{\partial}{\partial u}U_q(u)V_q(v) \quad h_v = \frac{1}{k_r l_v}\frac{\partial}{\partial v}U_q(u)V_q(v) \quad h_z = U_q(u)V_q(v) \ \textbf{(TE)} \tag{81}$$

The normalization is given by

$$\frac{1}{k_r^2}\int\left[\left(\frac{\partial}{\partial u}U_qV_q\right)\frac{l_v}{l_u}\left(\frac{\partial}{\partial u}U_qV_q\right)+\left(\frac{\partial}{\partial v}U_qV_q\right)\frac{l_u}{l_v}\left(\frac{\partial}{\partial v}U_qV_q\right)\right]dudv$$
$$=-\frac{1}{k_r^2}\int l_vl_uU_qV_q\Delta U_qV_q dudv=\int l_vl_uU_qV_qU_qV_q dudv=1 \quad . \tag{82}$$

Maxwell's equations with the original scaling are:

$$\left(\frac{1}{l_v}\frac{\partial}{\partial v}e_z-i\beta e_v\right)\mathbf{u}+\left(i\beta e_u-\frac{1}{l_u}\frac{\partial}{\partial u}e_z\right)\mathbf{v}+\frac{1}{l_ul_v}\left(\frac{\partial}{\partial u}l_ve_v-\frac{\partial}{\partial v}l_ue_u\right)\mathbf{z}=i\omega_0\mu_0\mathbf{h} \tag{83}$$

$$\left(\frac{1}{l_v}\frac{\partial}{\partial v}h_z-i\beta h_v\right)\mathbf{u}+\left(i\beta h_u-\frac{1}{l_u}\frac{\partial}{\partial u}h_z\right)\mathbf{v}+\frac{1}{l_ul_v}\left(\frac{\partial}{\partial u}l_vh_v-\frac{\partial}{\partial v}l_uh_u\right)\mathbf{z}=-i\omega_0\varepsilon_0\mathbf{e} \tag{84}$$

Rearranging gives:

$$i\left(\frac{1}{l_v}\frac{\partial}{\partial v}e_z\right)\mathbf{u}-i\left(\frac{1}{l_u}\frac{\partial}{\partial u}e_z\right)\mathbf{v}+\frac{i}{l_ul_v}\left(\frac{\partial}{\partial u}l_ve_v-\frac{\partial}{\partial v}l_ue_u\right)\mathbf{z}+\omega_0\mu_0\mathbf{h}=-\beta e_v\mathbf{u}+\beta e_u\mathbf{v} \tag{85}$$

$$i\left(\frac{1}{l_v}\frac{\partial}{\partial v}h_z\right)\mathbf{u}-i\left(\frac{1}{l_u}\frac{\partial}{\partial u}h_z\right)\mathbf{v}+\frac{i}{l_ul_v}\left(\frac{\partial}{\partial u}l_vh_v-\frac{\partial}{\partial v}l_uh_u\right)\mathbf{z}-\omega_0\varepsilon_0\mathbf{e}=-\beta h_v\mathbf{u}+\beta h_u\mathbf{v} \tag{86}$$

In matrix form this is:

$$\begin{bmatrix} 0 & 0 & i\frac{1}{l_v}\frac{\partial}{\partial v} & \omega_0\mu_0 & 0 & 0 \\ 0 & 0 & -i\frac{1}{l_u}\frac{\partial}{\partial u} & 0 & \omega_0\mu_0 & 0 \\ -i\frac{1}{l_ul_v}\frac{\partial}{\partial v}l_u & i\frac{1}{l_ul_v}\frac{\partial}{\partial u}l_v & 0 & 0 & 0 & \omega_0\mu_0 \\ -\omega_0\varepsilon & 0 & 0 & 0 & 0 & i\frac{1}{l_v}\frac{\partial}{\partial v} \\ 0 & -\omega_0\varepsilon & 0 & 0 & 0 & -i\frac{1}{l_u}\frac{\partial}{\partial u} \\ 0 & 0 & -\omega_0\varepsilon & -i\frac{1}{l_ul_v}\frac{\partial}{\partial v}l_u & i\frac{1}{l_ul_v}\frac{\partial}{\partial u}l_v & 0 \end{bmatrix}\begin{bmatrix} e_u \\ e_v \\ e_z \\ h_u \\ h_v \\ h_z \end{bmatrix}=\beta\begin{bmatrix} 0 & -1 & 0 & 0 & 0 & 0 \\ 1 & 0 & 0 & 0 & 0 & 0 \\ 0 & 0 & 0 & 0 & 0 & 0 \\ 0 & 0 & 0 & 0 & -1 & 0 \\ 0 & 0 & 0 & 1 & 0 & 0 \\ 0 & 0 & 0 & 0 & 0 & 0 \end{bmatrix}\begin{bmatrix} e_u \\ e_v \\ e_z \\ h_u \\ h_v \\ h_z \end{bmatrix} \tag{87}$$

This is modified slightly to make the RHS symmetric:

$$\begin{bmatrix} 0 & 0 & -i\frac{1}{l_u}\frac{\partial}{\partial u} & 0 & \omega_0\mu_0 & 0 \\ 0 & 0 & -i\frac{1}{l_v}\frac{\partial}{\partial v} & -\omega_0\mu_0 & 0 & 0 \\ -i\frac{1}{l_ul_v}\frac{\partial}{\partial v}l_u & i\frac{1}{l_ul_v}\frac{\partial}{\partial u}l_v & 0 & 0 & 0 & \omega_0\mu_0 \\ 0 & -\omega_0\varepsilon & 0 & 0 & 0 & -i\frac{1}{l_u}\frac{\partial}{\partial u} \\ \omega_0\varepsilon & 0 & 0 & 0 & 0 & -i\frac{1}{l_v}\frac{\partial}{\partial v} \\ 0 & 0 & -\omega_0\varepsilon & -i\frac{1}{l_ul_v}\frac{\partial}{\partial v}l_u & i\frac{1}{l_ul_v}\frac{\partial}{\partial u}l_v & 0 \end{bmatrix}\begin{bmatrix} e_u \\ e_v \\ e_z \\ h_u \\ h_v \\ h_z \end{bmatrix}=\beta\begin{bmatrix} 1 & 0 & 0 & 0 & 0 & 0 \\ 0 & 1 & 0 & 0 & 0 & 0 \\ 0 & 0 & 0 & 0 & 0 & 0 \\ 0 & 0 & 0 & 1 & 0 & 0 \\ 0 & 0 & 0 & 0 & 1 & 0 \\ 0 & 0 & 0 & 0 & 0 & 0 \end{bmatrix}\begin{bmatrix} e_u \\ e_v \\ e_z \\ h_u \\ h_v \\ h_z \end{bmatrix} \tag{88}$$

We can eliminate the $e_z$ and $h_z$ components using:

$$\frac{1}{\omega_0\mu_0}\frac{1}{l_u l_v}\frac{\partial}{\partial v}l_1 e_u - \frac{1}{\omega_0\mu_0}\frac{1}{l_u l_v}\frac{\partial}{\partial u}l_v e_v = -ih_z \tag{89}$$

$$-\frac{1}{\omega_0\varepsilon}\frac{1}{l_u l_v}\frac{\partial}{\partial v}l_u h_u + \frac{1}{\omega_0\varepsilon}\frac{1}{l_u l_v}\frac{\partial}{\partial u}l_v h_v = -ie_z \tag{90}$$

This gives:

$$\begin{bmatrix} 0 & 0 & -\frac{1}{l_u}\frac{\partial}{\partial u}\frac{1}{\omega_0\varepsilon}\frac{1}{l_u l_v}\frac{\partial}{\partial v}l_u & \omega_0\mu_0 + \frac{1}{l_u}\frac{\partial}{\partial u}\frac{1}{\omega_0\varepsilon}\frac{1}{l_u l_v}\frac{\partial}{\partial u}l_v \\ 0 & 0 & -\omega_0\mu_0 - \frac{1}{l_v}\frac{\partial}{\partial v}\frac{1}{\omega_0\varepsilon}\frac{1}{l_u l_v}\frac{\partial}{\partial v}l_u & \frac{1}{l_v}\frac{\partial}{\partial v}\frac{1}{\omega_0\varepsilon}\frac{1}{l_u l_v}\frac{\partial}{\partial u}l_v \\ \frac{1}{l_u}\frac{\partial}{\partial u}\frac{1}{\omega_0\mu_0}\frac{1}{l_u l_v}\frac{\partial}{\partial v}l_u & -\omega_0\varepsilon - \frac{1}{l_u}\frac{\partial}{\partial u}\frac{1}{\omega_0\mu_0}\frac{1}{l_u l_v}\frac{\partial}{\partial u}l_v & 0 & 0 \\ \omega_0\varepsilon + \frac{1}{l_v}\frac{\partial}{\partial v}\frac{1}{\omega_0\mu_0}\frac{1}{l_u l_v}\frac{\partial}{\partial v}l_u & -\frac{1}{l_v}\frac{\partial}{\partial v}\frac{1}{\omega_0\mu_0}\frac{1}{l_u l_v}\frac{\partial}{\partial u}l_v & 0 & 0 \end{bmatrix}\begin{bmatrix} e_u \\ e_v \\ h_u \\ h_v \end{bmatrix} \tag{91}$$
$$= \beta\begin{bmatrix} e_u \\ e_v \\ h_u \\ h_v \end{bmatrix}$$

The diagonal elements coupling the $e$ and $h$ field components with themselves are all zero. The above can thus be written as:

$$\hat{O}_{eh}\begin{bmatrix} h_u \\ h_v \end{bmatrix} = \beta\begin{bmatrix} e_u \\ e_v \end{bmatrix} \tag{92}$$

$$\hat{O}_{he}\begin{bmatrix} e_u \\ e_v \end{bmatrix} = \beta\begin{bmatrix} h_u \\ h_v \end{bmatrix} \tag{93}$$

Reintroducing the scaling $\mathbf{H}\omega_0\mu_0 = \mathbf{H}'$, the $\hat{O}_{eh}$ and $\hat{O}_{he}$ operator are

$$\hat{O}_{eh} = \begin{bmatrix} -\frac{1}{l_u}\frac{\partial}{\partial u}\frac{1}{k_0^2\varepsilon_r}\frac{1}{l_u l_v}\frac{\partial}{\partial v}l_u & 1 + \frac{1}{l_u}\frac{\partial}{\partial u}\frac{1}{k_0^2\varepsilon_r}\frac{1}{l_u l_v}\frac{\partial}{\partial u}l_v \\ -1 - \frac{1}{l_v}\frac{\partial}{\partial v}\frac{1}{k_0^2\varepsilon_r}\frac{1}{l_u l_v}\frac{\partial}{\partial v}l_u & \frac{1}{l_v}\frac{\partial}{\partial v}\frac{1}{k_0^2\varepsilon_r}\frac{1}{l_u l_v}\frac{\partial}{\partial u}l_v \end{bmatrix} \tag{94}$$

$$\hat{O}_{he} = \begin{bmatrix} \frac{1}{l_u}\frac{\partial}{\partial u}\frac{1}{l_u l_v}\frac{\partial}{\partial v}l_u & -k_0^2\varepsilon_r - \frac{1}{l_u}\frac{\partial}{\partial u}\frac{1}{l_u l_v}\frac{\partial}{\partial u}l_v \\ k_0^2\varepsilon_r + \frac{1}{l_v}\frac{\partial}{\partial v}\frac{1}{l_u l_v}\frac{\partial}{\partial v}l_u & -\frac{1}{l_v}\frac{\partial}{\partial v}\frac{1}{l_u l_v}\frac{\partial}{\partial u}l_v \end{bmatrix}. \tag{95}$$

To implement explicit support for emulating open boundary conditions using a coordinate transformation, we introduce the transformation

$$l_u \to \frac{l_u}{F(u)}\ , \tag{96}$$

where $F(u)$ represents a path into the complex plane. The operators (94)-(95) then become

$$\hat{O}_{eh}=\begin{bmatrix} -\frac{F}{l_u}\frac{\partial}{\partial u}\frac{1}{k_0^2\varepsilon_r}\frac{F}{l_ul_v}\frac{\partial}{\partial v}l_u & 1+\frac{F}{l_u}\frac{\partial}{\partial u}\frac{1}{k_0^2\varepsilon_r}\frac{F}{l_ul_v}\frac{\partial}{\partial u}l_v \\ -1-\frac{F}{l_v}\frac{\partial}{\partial v}\frac{1}{k_0^2\varepsilon_r}\frac{F}{l_ul_v}\frac{\partial}{\partial v}l_u & \frac{F}{l_v}\frac{\partial}{\partial v}\frac{1}{k_0^2\varepsilon_r}\frac{F}{l_ul_v}\frac{\partial}{\partial u}l_v \end{bmatrix} \quad (97)$$

$$\hat{O}_{he}=\begin{bmatrix} \frac{F}{l_u}\frac{\partial}{\partial u}\frac{F}{l_ul_v}\frac{\partial}{\partial v}l_u & -k_0^2\varepsilon_r-\frac{F}{l_u}\frac{\partial}{\partial u}\frac{F}{l_ul_v}\frac{\partial}{\partial u}l_v \\ k_0^2\varepsilon_r+\frac{F}{l_v}\frac{\partial}{\partial v}\frac{F}{l_ul_v}\frac{\partial}{\partial v}l_u & -\frac{F}{l_v}\frac{\partial}{\partial v}\frac{F}{l_ul_v}\frac{\partial}{\partial u}l_v \end{bmatrix}. \quad (98)$$

In the following we write the regular operators as function of TM and TE contributions. The TM-TM contribution is given by

$$\left\langle \mathbf{e}_{TM}\middle|\hat{O}_{eh}\middle|\mathbf{h}_{TM}\right\rangle=\frac{1}{k_{rL}}\int l_ul_v\begin{bmatrix}\frac{1}{l_u}\frac{\partial}{\partial u}U_q(u)V_q(v) & \frac{1}{l_v}\frac{\partial}{\partial v}U_q(u)V_q(v)\end{bmatrix}$$

$$\begin{bmatrix} -\frac{F}{l_u}\frac{\partial}{\partial u}\frac{1}{k_0^2\varepsilon_r}\frac{F}{l_ul_v}\frac{\partial}{\partial v}l_u & 1+\frac{F}{l_u}\frac{\partial}{\partial u}\frac{1}{k_0^2\varepsilon_r}\frac{F}{l_ul_v}\frac{\partial}{\partial u}l_v \\ -1-\frac{F}{l_v}\frac{\partial}{\partial v}\frac{1}{k_0^2\varepsilon_r}\frac{F}{l_ul_v}\frac{\partial}{\partial v}l_u & \frac{F}{l_v}\frac{\partial}{\partial v}\frac{1}{k_0^2\varepsilon_r}\frac{F}{l_ul_v}\frac{\partial}{\partial u}l_v \end{bmatrix}\frac{1}{k_{rR}}\begin{bmatrix}-\frac{1}{l_v}\frac{\partial}{\partial v} \\ \frac{1}{l_u}\frac{\partial}{\partial u}\end{bmatrix}U_q(u)V_q(v)dudv \quad . \quad (99)$$

This becomes

$$\frac{1}{k_{rL}}\int l_ul_v\begin{bmatrix}\frac{1}{l_u}\frac{\partial}{\partial u}U_q(u)V_q(v) & \frac{1}{l_v}\frac{\partial}{\partial v}U_q(u)V_q(v)\end{bmatrix}$$

$$\begin{bmatrix} \frac{F}{l_u}\frac{\partial}{\partial u}\frac{1}{k_0^2\varepsilon_r}F\Delta\frac{1}{k_{rR}}+\frac{1}{k_{rR}l_u}\frac{\partial}{\partial u} \\ \frac{1}{k_{rR}l_v}\frac{\partial}{\partial v}+\frac{F}{l_v}\frac{\partial}{\partial v}\frac{1}{k_0^2\varepsilon_r}F\Delta\frac{1}{k_{rR}} \end{bmatrix}U_q(u)V_q(v)dudv, \quad (100)$$

which equals

$$\frac{1}{k_{rL}}\int\begin{bmatrix} \left(\frac{\partial}{\partial u}U_qV_q\right)\frac{l_v}{l_u}\left[\frac{\partial}{\partial u}\frac{1}{k_{rR}}+F\frac{\partial}{\partial u}\frac{F}{k_0^2\varepsilon_r}\Delta\right]U_qV_q \\ +\left(\frac{\partial}{\partial v}U_qV_q\right)\frac{l_u}{l_v}\left[\frac{\partial}{\partial v}\frac{1}{k_{rR}}+F\frac{\partial}{\partial v}\frac{F}{k_0^2\varepsilon_r}\Delta\right]U_qV_q \end{bmatrix}\frac{1}{k_{rR}}dudv$$

$$=-\frac{1}{k_{rL}}\int U_qV_q\left(\frac{\partial}{\partial u}\frac{l_v}{l_u}F\frac{\partial}{\partial u}+\frac{\partial}{\partial v}\frac{l_u}{l_v}F\frac{\partial}{\partial v}\right)\frac{F}{k_0^2\varepsilon_r}\Delta U_qV_q\frac{1}{k_{rR}}dudv$$

$$-\frac{1}{k_{rL}}\int l_ul_vU_qV_q\Delta U_qV_q\frac{1}{k_{rR}}dudv \quad . \quad (101)$$

$$=-\frac{1}{k_{rL}}\int U_qV_q\left(\frac{\partial}{\partial u}F\right)\frac{l_v}{l_u}\frac{\partial}{\partial u}\frac{F}{k_0^2\varepsilon_r}\Delta U_qV_q\frac{1}{k_{rR}}dudv$$

$$-\frac{1}{k_{rL}}\int l_ul_vU_qV_qF\Delta\frac{F}{k_0^2\varepsilon_r}\Delta\frac{1}{k_{rR}}U_qV_qdudv+\delta_{mn}$$

The TE-TE contribution is

$$\left\langle \mathbf{e}_{TE} \left| \hat{O}_{eh} \right| \mathbf{h}_{TE} \right\rangle = \frac{1}{k_{rL}} \int l_u l_v \begin{bmatrix} \frac{1}{l_v}\frac{\partial}{\partial v}U_q(u)V_q(v) & -\frac{1}{l_u}\frac{\partial}{\partial u}U_q(u)V_q(v) \end{bmatrix}$$
$$\begin{bmatrix} -\frac{F}{l_u}\frac{\partial}{\partial u}\frac{1}{k_0^2\varepsilon_r}\frac{F}{l_u l_v}\frac{\partial}{\partial v}l_u & 1+\frac{F}{l_u}\frac{\partial}{\partial u}\frac{1}{k_0^2\varepsilon_r}\frac{F}{l_u l_v}\frac{\partial}{\partial u}l_v \\ -1-\frac{F}{l_v}\frac{\partial}{\partial v}\frac{1}{k_0^2\varepsilon_r}\frac{F}{l_u l_v}\frac{\partial}{\partial v}l_u & \frac{F}{l_v}\frac{\partial}{\partial v}\frac{1}{k_0^2\varepsilon_r}\frac{F}{l_u l_v}\frac{\partial}{\partial u}l_v \end{bmatrix} \frac{1}{k_{rR}} \begin{bmatrix} \frac{1}{l_u}\frac{\partial}{\partial u}U_q(u)V_q(v) \\ \frac{1}{l_v}\frac{\partial}{\partial v}U_q(u)V_q(v) \end{bmatrix} dudv \quad . \tag{102}$$

This becomes

$$\frac{1}{k_{rL}} \int \begin{bmatrix} \frac{1}{l_v}\frac{\partial}{\partial v}U_q(u)V_q(v) & -\frac{1}{l_u}\frac{\partial}{\partial u}U_q(u)V_q(v) \end{bmatrix}$$
$$\frac{1}{k_{rR}} \begin{bmatrix} \frac{1}{l_v}\frac{\partial}{\partial v}U_q(u)V_q(v) \\ -\frac{1}{l_u}\frac{\partial}{\partial u}U_q(u)V_q(v) \end{bmatrix} l_u l_v dudv = \frac{1}{k_{rL}k_{rR}} \int l_u l_v U_q V_q \Delta U_q V_q dudv = \delta_{mn} \quad . \tag{103}$$

The TM-TE contribution is

$$\left\langle \mathbf{e}_{TM} \left| \hat{O}_{eh} \right| \mathbf{h}_{TE} \right\rangle = \frac{1}{k_{rL}} \int l_u l_v \begin{bmatrix} \frac{1}{l_u}\frac{\partial}{\partial u}U_q(u)V_q(v) & \frac{1}{l_v}\frac{\partial}{\partial v}U_q(u)V_q(v) \end{bmatrix}$$
$$\begin{bmatrix} -\frac{F}{l_u}\frac{\partial}{\partial u}\frac{1}{k_0^2\varepsilon_r}\frac{F}{l_u l_v}\frac{\partial}{\partial v}l_u & 1+\frac{F}{l_u}\frac{\partial}{\partial u}\frac{1}{k_0^2\varepsilon_r}\frac{F}{l_u l_v}\frac{\partial}{\partial u}l_v \\ -1-\frac{F}{l_v}\frac{\partial}{\partial v}\frac{1}{k_0^2\varepsilon_r}\frac{F}{l_u l_v}\frac{\partial}{\partial v}l_u & \frac{F}{l_v}\frac{\partial}{\partial v}\frac{1}{k_0^2\varepsilon_r}\frac{F}{l_u l_v}\frac{\partial}{\partial u}l_v \end{bmatrix} \frac{1}{k_{rR}} \begin{bmatrix} \frac{1}{l_u}\frac{\partial}{\partial u}U_q(u)V_q(v) \\ \frac{1}{l_v}\frac{\partial}{\partial v}U_q(u)V_q(v) \end{bmatrix} dudv \quad . \tag{104}$$

This becomes

$$\frac{1}{k_{rL}k_{rR}} \int \begin{bmatrix} \frac{1}{l_u}\frac{\partial}{\partial u}U_q(u)V_q(v) & \frac{1}{l_v}\frac{\partial}{\partial v}U_q(u)V_q(v) \end{bmatrix} \begin{bmatrix} \frac{1}{l_v}\frac{\partial}{\partial v}U_q(u)V_q(v) \\ -\frac{1}{l_u}\frac{\partial}{\partial u}U_q(u)V_q(v) \end{bmatrix} l_u l_v dudv = 0 \,. \tag{105}$$

The TE-TM contribution is

$$\left\langle \mathbf{e}_{TE} \left| \hat{O}_{eh} \right| \mathbf{h}_{TM} \right\rangle = \frac{1}{k_{rL}} \int l_u l_v \begin{bmatrix} \frac{1}{l_v}\frac{\partial}{\partial v}U_q(u)V_q(v) & -\frac{1}{l_u}\frac{\partial}{\partial u}U_q(u)V_q(v) \end{bmatrix}$$
$$\begin{bmatrix} -\frac{F}{l_u}\frac{\partial}{\partial u}\frac{1}{k_0^2\varepsilon_r}\frac{F}{l_u l_v}\frac{\partial}{\partial v}l_u & 1+\frac{F}{l_u}\frac{\partial}{\partial u}\frac{1}{k_0^2\varepsilon_r}\frac{F}{l_u l_v}\frac{\partial}{\partial u}l_v \\ -1-\frac{F}{l_v}\frac{\partial}{\partial v}\frac{1}{k_0^2\varepsilon_r}\frac{F}{l_u l_v}\frac{\partial}{\partial v}l_u & \frac{F}{l_v}\frac{\partial}{\partial v}\frac{1}{k_0^2\varepsilon_r}\frac{F}{l_u l_v}\frac{\partial}{\partial u}l_v \end{bmatrix} \frac{1}{k_{rR}} \begin{bmatrix} -\frac{1}{l_v}\frac{\partial}{\partial v} \\ \frac{1}{l_u}\frac{\partial}{\partial u} \end{bmatrix} U_q(u)V_q(v) dudv \quad . \tag{106}$$

This becomes

$$\frac{1}{k_{rL}}\int l_u l_v \left[\frac{1}{l_v}\frac{\partial}{\partial v}U_q(u)V_q(v) \quad -\frac{1}{l_u}\frac{\partial}{\partial u}U_q(u)V_q(v)\right]$$
$$\begin{bmatrix} \frac{F}{l_u}\frac{\partial}{\partial u}\frac{F}{k_0^2\varepsilon_r}\Delta+\frac{1}{l_u}\frac{\partial}{\partial u} \\ \frac{1}{l_v}\frac{\partial}{\partial v}+\frac{F}{l_v}\frac{\partial}{\partial v}\frac{F}{k_0^2\varepsilon_r}\Delta \end{bmatrix}\frac{1}{k_{rR}}U_q(u)V_q(v)dudv \quad . \tag{107}$$
$$=\frac{1}{k_{rL}}\int\left[\left(\frac{\partial}{\partial v}U_qV_q\right)F\frac{\partial}{\partial u}-\left(\frac{\partial}{\partial u}U_qV_q\right)F\frac{\partial}{\partial v}\right]\frac{F}{k_0^2\varepsilon_r}\Delta U_qV_q\frac{1}{k_{rR}}dudv$$
$$=\frac{1}{k_{rL}}\int U_qV_q\left(\frac{\partial}{\partial u}F\right)\frac{\partial}{\partial v}\frac{F}{k_0^2\varepsilon_r}\Delta U_qV_q\frac{1}{k_{rR}}dudv$$

The TM-TM contribution for the $\hat{O}_{he}$ operator is given by

$$\left\langle \mathbf{h}_{TM}\left|\hat{O}_{he}\right|\mathbf{e}_{TM}\right\rangle=\frac{1}{k_{rL}}\int l_u l_v\left[-\frac{1}{l_v}\frac{\partial}{\partial v}U_q(u)V_q(v) \quad \frac{1}{l_u}\frac{\partial}{\partial u}U_q(u)V_q(v)\right]$$
$$\begin{bmatrix} \frac{F}{l_u}\frac{\partial}{\partial u}\frac{F}{l_ul_v}\frac{\partial}{\partial v}l_u & -k_0^2\varepsilon_r-\frac{F}{l_u}\frac{\partial}{\partial u}\frac{F}{l_ul_v}\frac{\partial}{\partial u}l_v \\ k_0^2\varepsilon_r+\frac{F}{l_v}\frac{\partial}{\partial v}\frac{F}{l_ul_v}\frac{\partial}{\partial v}l_u & -\frac{F}{l_v}\frac{\partial}{\partial v}\frac{F}{l_ul_v}\frac{\partial}{\partial u}l_v \end{bmatrix}\frac{1}{k_{rR}}\begin{bmatrix}\frac{1}{l_u}\frac{\partial}{\partial u} \\ \frac{1}{l_v}\frac{\partial}{\partial v}\end{bmatrix}U_q(u)V_q(v)dudv \quad . \tag{108}$$

This becomes

$$\frac{1}{k_{rL}}\int l_u l_v\left[-\frac{1}{l_v}\frac{\partial}{\partial v}U_q(u)V_q(v) \quad \frac{1}{l_u}\frac{\partial}{\partial u}U_q(u)V_q(v)\right]$$
$$\frac{1}{k_{rR}}\begin{bmatrix} -k_0^2\varepsilon_r\frac{1}{l_v}\frac{\partial}{\partial v} \\ k_0^2\varepsilon_r\frac{1}{l_u}\frac{\partial}{\partial u}\end{bmatrix}U_q(u)V_q(v)dudv \quad , \tag{109}$$

which equals

$$\frac{1}{k_{rL}}\int\left[\left(\frac{\partial}{\partial u}U_qV_q\right)k_0^2\varepsilon_r\frac{l_v}{l_u}\left(\frac{\partial}{\partial u}U_qV_q\right)+\left(\frac{\partial}{\partial v}U_qV_q\right)k_0^2\varepsilon_r\frac{l_u}{l_v}\left(\frac{\partial}{\partial v}U_qV_q\right)\right]\frac{1}{k_{rR}}dudv$$
$$=-\frac{1}{k_{rL}}\int l_ul_vU_qV_qk_0^2\varepsilon_r\Delta U_qV_q\frac{1}{k_{rR}}dudv-\frac{1}{k_{rL}}\int U_qV_qk_0^2\left(\frac{\partial}{\partial u}\varepsilon_r\right)\frac{l_v}{l_u}\left(\frac{\partial}{\partial u}U_q\right)V_q\frac{1}{k_{rR}}dudv \quad . \tag{110}$$

The TE-TE contribution is

$$\left\langle \mathbf{h}_{TE}\left|\hat{O}_{he}\right|\mathbf{e}_{TE}\right\rangle=\frac{1}{k_{rL}}\int\left[\frac{1}{l_u}\frac{\partial}{\partial u}U_q(u)V_q(v)\quad \frac{1}{l_v}\frac{\partial}{\partial v}U_q(u)V_q(v)\right]$$

$$\begin{bmatrix}\frac{F}{l_u}\frac{\partial}{\partial u}\frac{F}{l_ul_v}\frac{\partial}{\partial v}l_u & -k_0^2\varepsilon_r-\frac{F}{l_u}\frac{\partial}{\partial u}\frac{F}{l_ul_v}\frac{\partial}{\partial u}l_v\\ k_0^2\varepsilon_r+\frac{F}{l_v}\frac{\partial}{\partial v}\frac{F}{l_ul_v}\frac{\partial}{\partial v}l_u & -\frac{F}{l_v}\frac{\partial}{\partial v}\frac{F}{l_ul_v}\frac{\partial}{\partial u}l_v\end{bmatrix}\frac{1}{k_{rR}}\begin{bmatrix}\frac{1}{l_v}\frac{\partial}{\partial v}\\ -\frac{1}{l_u}\frac{\partial}{\partial u}\end{bmatrix}U_q(u)V_q(v)l_ul_v dudv \quad . \tag{111}$$

This becomes

$$\frac{1}{k_{rL}}\int\left[\frac{1}{l_u}\frac{\partial}{\partial u}U_q(u)V_q(v)\quad \frac{1}{l_v}\frac{\partial}{\partial v}U_q(u)V_q(v)\right]$$

$$\begin{bmatrix}\frac{1}{l_u}F\frac{\partial}{\partial u}\frac{1}{l_ul_v}F\left(\frac{\partial}{\partial v}\frac{l_u}{l_v}\frac{\partial}{\partial v}+\frac{\partial}{\partial u}\frac{l_v}{l_u}\frac{\partial}{\partial u}\right)+k_0^2\varepsilon_r\frac{1}{l_u}\frac{\partial}{\partial u}\\ \frac{1}{l_v}F\frac{\partial}{\partial v}\frac{1}{l_ul_v}F\left(\frac{\partial}{\partial v}\frac{l_u}{l_v}\frac{\partial}{\partial v}+\frac{\partial}{\partial u}\frac{l_v}{l_u}\frac{\partial}{\partial u}\right)+k_0^2\varepsilon_r\frac{1}{l_v}\frac{\partial}{\partial v}\end{bmatrix}\frac{1}{k_{rR}}U_q(u)V_q(v)l_ul_v dudv \quad , \tag{112}$$

which equals

$$\frac{1}{k_{rL}}\int\begin{bmatrix}\left(\frac{\partial}{\partial u}U_qV_q\right)\frac{l_v}{l_u}\left(k_0^2\varepsilon_r\frac{\partial}{\partial u}U_qV_q+F\frac{\partial}{\partial u}F\Delta U_qV_q\right)\\ +\left(\frac{\partial}{\partial v}U_qV_q\right)\frac{l_u}{l_v}\left(k_0^2\varepsilon_r\frac{\partial}{\partial v}U_qV_q+F\frac{\partial}{\partial v}F\Delta U_qV_q\right)\end{bmatrix}dudv\frac{1}{k_{rR}}$$

$$\begin{aligned}
&=-\frac{1}{k_{rL}}\int U_qV_q\left(\frac{\partial}{\partial u}\frac{l_v}{l_u}F\frac{\partial}{\partial u}+\frac{\partial}{\partial v}\frac{l_u}{l_v}F\frac{\partial}{\partial v}\right)F\Delta U_qV_q\frac{1}{k_{rR}}dudv\\
&-\frac{1}{k_{rL}}\int l_ul_vU_qV_qk_0^2\varepsilon_r\Delta U_qV_q\frac{1}{k_{rR}}dudv-\frac{1}{k_{rL}}\int U_qV_qk_0^2\left(\frac{\partial}{\partial u}\varepsilon_r\right)\frac{l_v}{l_u}\left(\frac{\partial}{\partial u}U_q\right)V_q\frac{1}{k_{rR}}dudv \cdot\\
&=-\frac{1}{k_{rL}}\int U_qV_q\left(\frac{\partial}{\partial u}F\right)\frac{l_v}{l_u}\frac{\partial}{\partial u}F\Delta U_qV_q\frac{1}{k_{rR}}dudv\\
&-\frac{1}{k_{rL}}\int l_ul_vU_qV_qF\Delta F\Delta\frac{1}{k_{rR}}U_qV_qdudv\\
&-\frac{1}{k_{rL}}\int l_ul_vU_qV_qk_0^2\varepsilon_r\Delta U_qV_q\frac{1}{k_{rR}}dudv-\frac{1}{k_{rL}}\int U_qV_qk_0^2\left(\frac{\partial}{\partial u}\varepsilon_r\right)\frac{l_v}{l_u}\left(\frac{\partial}{\partial u}U_q\right)V_q\frac{1}{k_{rR}}dudv
\end{aligned} \tag{113}$$

The TM-TE contribution is

$$\left\langle \mathbf{h}_{TM}\left|\hat{O}_{he}\right|\mathbf{e}_{TE}\right\rangle=\frac{1}{k_{rL}}\int\left[-\frac{1}{l_v}\frac{\partial}{\partial v}U_q(u)V_q(v)\quad \frac{1}{l_u}\frac{\partial}{\partial u}U_q(u)V_q(v)\right]$$

$$\begin{bmatrix}\frac{F}{l_u}\frac{\partial}{\partial u}\frac{F}{l_ul_v}\frac{\partial}{\partial v}l_u & -k_0^2\varepsilon_r-\frac{F}{l_u}\frac{\partial}{\partial u}\frac{F}{l_ul_v}\frac{\partial}{\partial u}l_v\\ k_0^2\varepsilon_r+\frac{F}{l_v}\frac{\partial}{\partial v}\frac{F}{l_ul_v}\frac{\partial}{\partial v}l_u & -\frac{F}{l_v}\frac{\partial}{\partial v}\frac{F}{l_ul_v}\frac{\partial}{\partial u}l_v\end{bmatrix}\begin{bmatrix}\frac{1}{l_v}\frac{\partial}{\partial v}\\ -\frac{1}{l_u}\frac{\partial}{\partial u}\end{bmatrix}U_q(u)V_q(v)l_ul_v\frac{1}{k_{rR}}dudv \quad . \tag{114}$$

This becomes

$$\frac{1}{k_{rL}}\int\left[-\frac{1}{l_v}\frac{\partial}{\partial v}U_q(u)V_q(v) \quad \frac{1}{l_u}\frac{\partial}{\partial u}U_q(u)V_q(v)\right]$$
$$\begin{bmatrix}\frac{F}{l_u}\frac{\partial}{\partial u}F\Delta+k_0^2\varepsilon_r\frac{1}{l_u}\frac{\partial}{\partial u}\\ \frac{F}{l_v}\frac{\partial}{\partial v}F\Delta+k_0^2\varepsilon_r\frac{1}{l_v}\frac{\partial}{\partial v}\end{bmatrix}U_q(u)V_q(v)l_ul_vdudv\frac{1}{k_{rR}}, \tag{115}$$

which equals

$$\begin{aligned}&\frac{1}{k_{rL}}\int\left[-\left(\frac{\partial}{\partial v}U_qV_q\right)F\frac{\partial}{\partial u}+\left(\frac{\partial}{\partial u}U_qV_q\right)F\frac{\partial}{\partial v}\right]F\Delta U_qV_q\frac{1}{k_{rR}}dudv\\&+\frac{1}{k_{rL}}\int\left[-\left(\frac{\partial}{\partial v}U_qV_q\right)k_0^2\varepsilon_r\left(\frac{\partial}{\partial u}U_qV_q\right)+\left(\frac{\partial}{\partial u}U_qV_q\right)k_0^2\varepsilon_r\left(\frac{\partial}{\partial v}U_qV_q\right)\right]\frac{1}{k_{rR}}dudv\\&=-\frac{1}{k_{rL}}\int U_qV_q\left(\frac{\partial}{\partial u}F\right)\frac{\partial}{\partial v}F\Delta U_qV_q\frac{1}{k_{rR}}dudv\\&-\frac{1}{k_{rL}}\int U_qV_qk_0^2\left(\frac{\partial}{\partial u}\varepsilon_r\right)\left(\frac{\partial}{\partial v}U_qV_q\right)\frac{1}{k_{rR}}dudv\end{aligned}. \tag{116}$$

The TE-TM contribution is

$$\left\langle\mathbf{h}_{TE}\left|\hat{O}_{he}\right|\mathbf{e}_{TM}\right\rangle=\frac{1}{k_{rL}}\int\left[\frac{1}{l_u}\frac{\partial}{\partial u}U_q(u)V_q(v) \quad \frac{1}{l_v}\frac{\partial}{\partial v}U_q(u)V_q(v)\right]$$
$$\begin{bmatrix}\frac{F}{l_u}\frac{\partial}{\partial u}\frac{F}{l_ul_v}\frac{\partial}{\partial v}l_u & -k_0^2\varepsilon_r-\frac{F}{l_u}\frac{\partial}{\partial u}\frac{F}{l_ul_v}\frac{\partial}{\partial u}l_v\\ k_0^2\varepsilon_r+\frac{F}{l_v}\frac{\partial}{\partial v}\frac{F}{l_ul_v}\frac{\partial}{\partial v}l_u & -\frac{F}{l_v}\frac{\partial}{\partial v}\frac{F}{l_ul_v}\frac{\partial}{\partial u}l_v\end{bmatrix}\begin{bmatrix}\frac{1}{l_u}\frac{\partial}{\partial u}\\ \frac{1}{l_v}\frac{\partial}{\partial v}\end{bmatrix}U_q(u)V_q(v)l_ul_vdudv\frac{1}{k_{rR}}. \tag{117}$$

This becomes

$$\frac{1}{k_{rL}}\int\left[\frac{1}{l_u}\frac{\partial}{\partial u}U_q(u)V_q(v) \quad \frac{1}{l_v}\frac{\partial}{\partial v}U_q(u)V_q(v)\right]$$
$$\begin{bmatrix}-k_0^2\varepsilon_r\frac{1}{l_v}\frac{\partial}{\partial v}\\ k_0^2\varepsilon_r\frac{1}{l_u}\frac{\partial}{\partial u}\end{bmatrix}U_q(u)V_q(v)l_ul_vdudv\frac{1}{k_{rR}}, \tag{118}$$

which equals

$$\begin{aligned}&\frac{1}{k_{rL}}\int\left[-\left(\frac{\partial}{\partial u}U_qV_q\right)k_0^2\varepsilon_r\left(\frac{\partial}{\partial v}U_qV_q\right)+\left(\frac{\partial}{\partial v}U_qV_q\right)k_0^2\varepsilon_r\left(\frac{\partial}{\partial u}U_qV_q\right)\right]\frac{1}{k_{rR}}dudv\\&=\frac{1}{k_{rL}}\int U_qV_qk_0^2\left(\frac{\partial}{\partial u}\varepsilon_r\right)\left(\frac{\partial}{\partial v}U_qV_q\right)\frac{1}{k_{rR}}dudv\end{aligned}. \tag{119}$$

# 4 Photonic NW design implementing an elliptical cross section

In the following we employ the new method to identify the ideal geometrical parameters for the bottom metal mirror section [10] and to study the influence of the top annular gold ring on the transmission through the top contact section [8].

## 4.1 Bottom metal mirror section

The bottom part of the elliptical photonic nanowire geometry is illustrated in Fig. 2. The ideal cross-section of the elliptical nanowire for polarization control has been established previously [2]. Our objective is now to compute the optimal distance *h* between the quantum dot (QD) and the bottom metal mirror as well as the optimal thickness *t* of the silica layer between the GaAs nanowire and the gold. Correct vertical positioning of the QD is crucial for achieving an efficient coupling to the fundamental $HE_{11}$ waveguide mode.

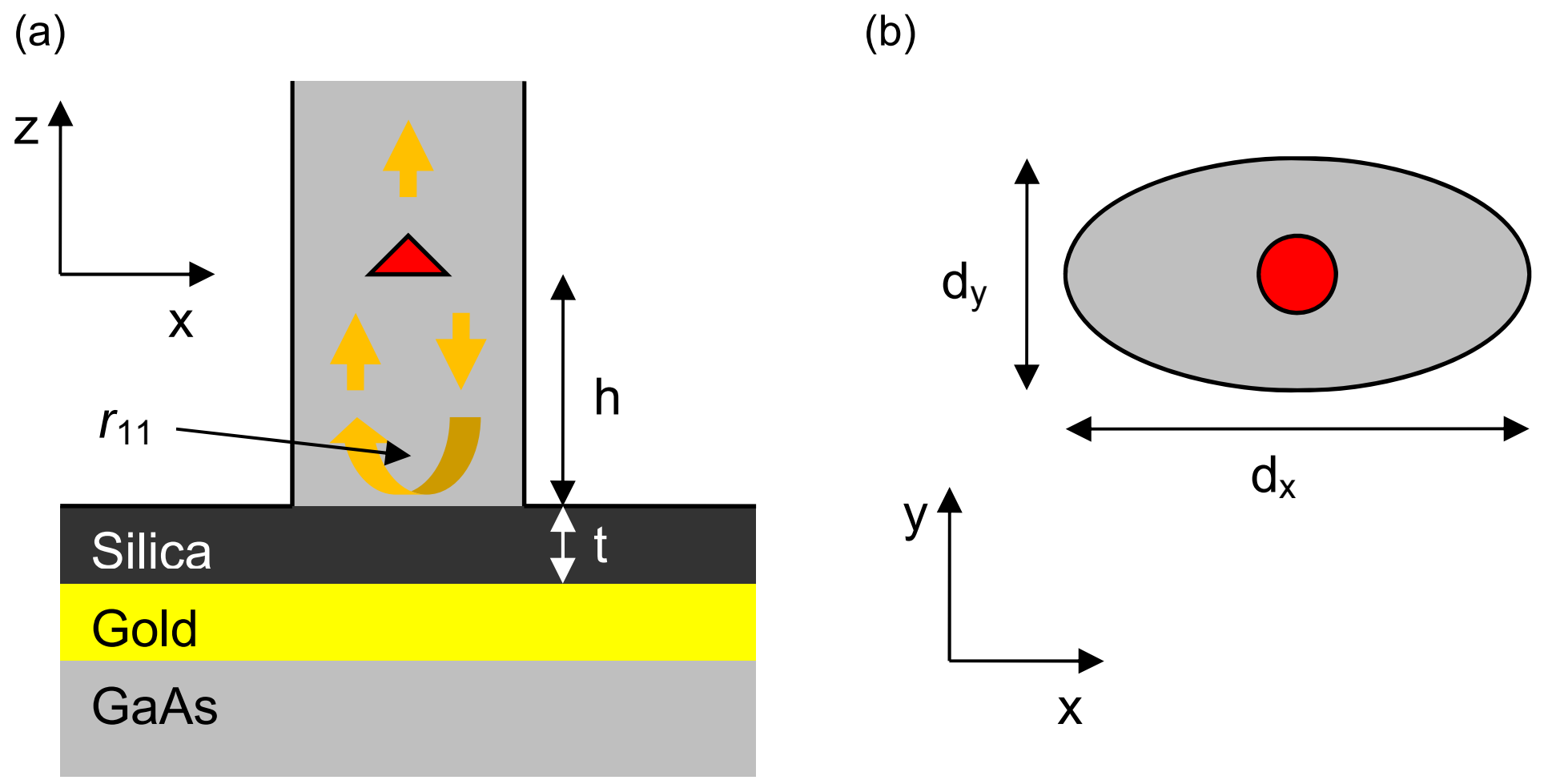

**Fig. 2. (a) Bottom part of the elliptical nanowire geometry. (b) Cross-section.**

In the following, we compute the reflection $R = |r_{11}|^2$ and the ideal QD-mirror distance *h* as function of the thickness *t* of the silica layer for the ideal cross-section parameters $d_x$ = 280 nm and $d_y$ = 130 nm.

To verify the calculations, convergence checks are performed as function of the number of included modes per order and included angular orders $O_{max}$ in the calculation. Furthermore, a convergence check with respect to the radius $R_{CD}$ (along the short semi-axis) of the computational domain is performed.

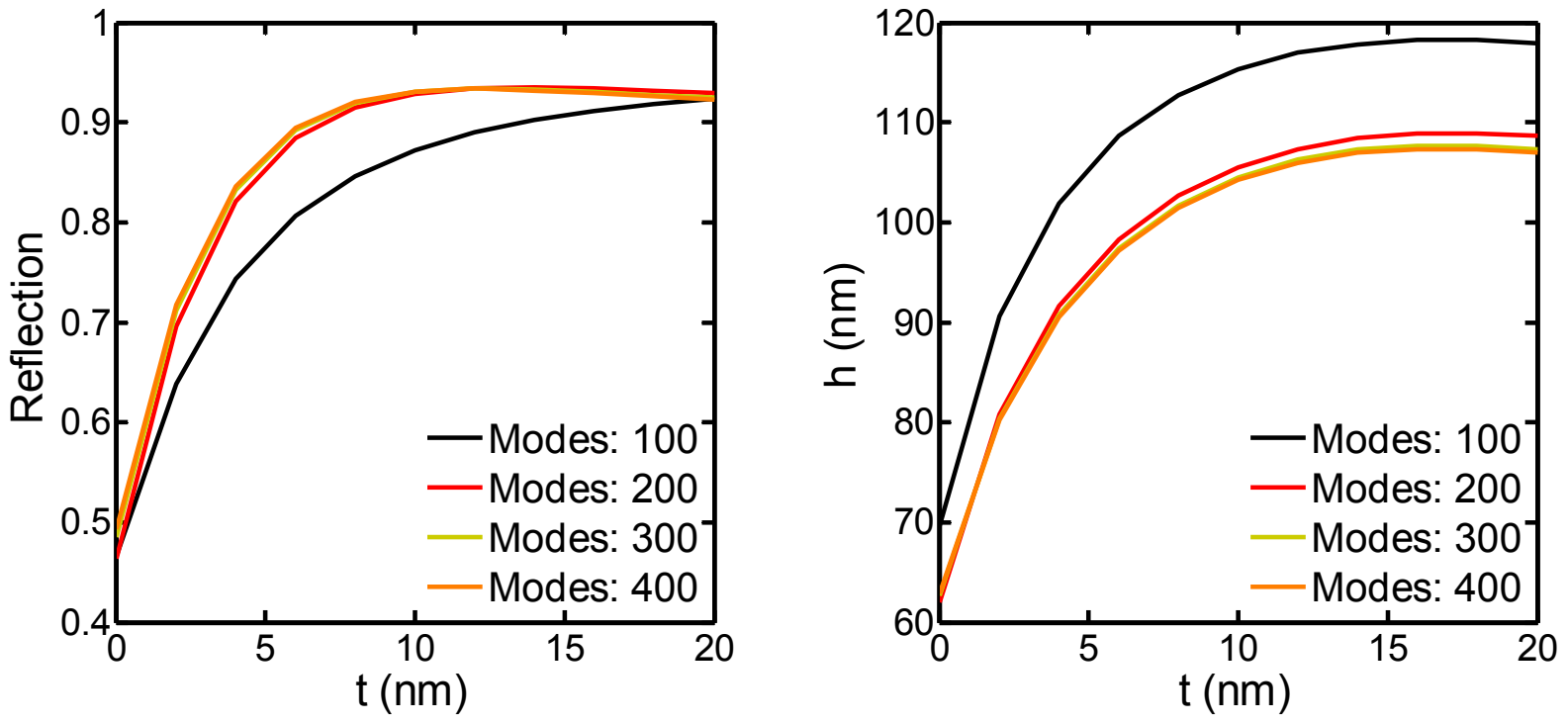

**Fig. 3. Reflection (left) and ideal QD-mirror distance *h* (right) as function of the silica layer thickness *t* and the number of modes per order. Parameters: $O_{max}$: 2 and $R_{CD}$: 3 μm.**

The number of required modes is studied in Fig. 3. We observe that convergence is obtained for approximately 400 modes.

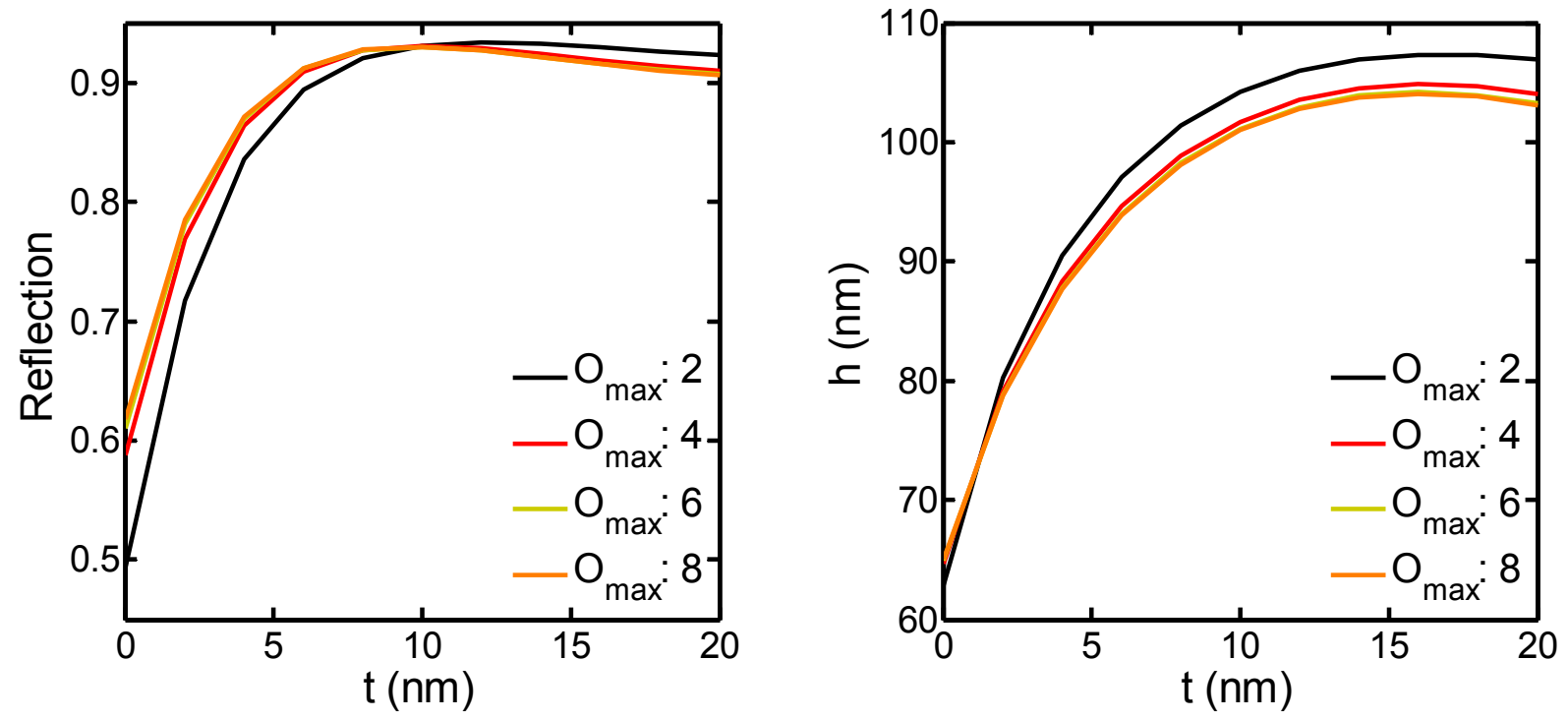

**Fig. 4. Reflection and *h* as function of *t* and the number of included orders $O_{max}$. Parameters: Modes: 400 and $R_{CD}$: 3 μm.**

The number of required orders is studied in Fig. 4. We observe that convergence is obtained for approximately $O_{max}$ = 6 included orders.

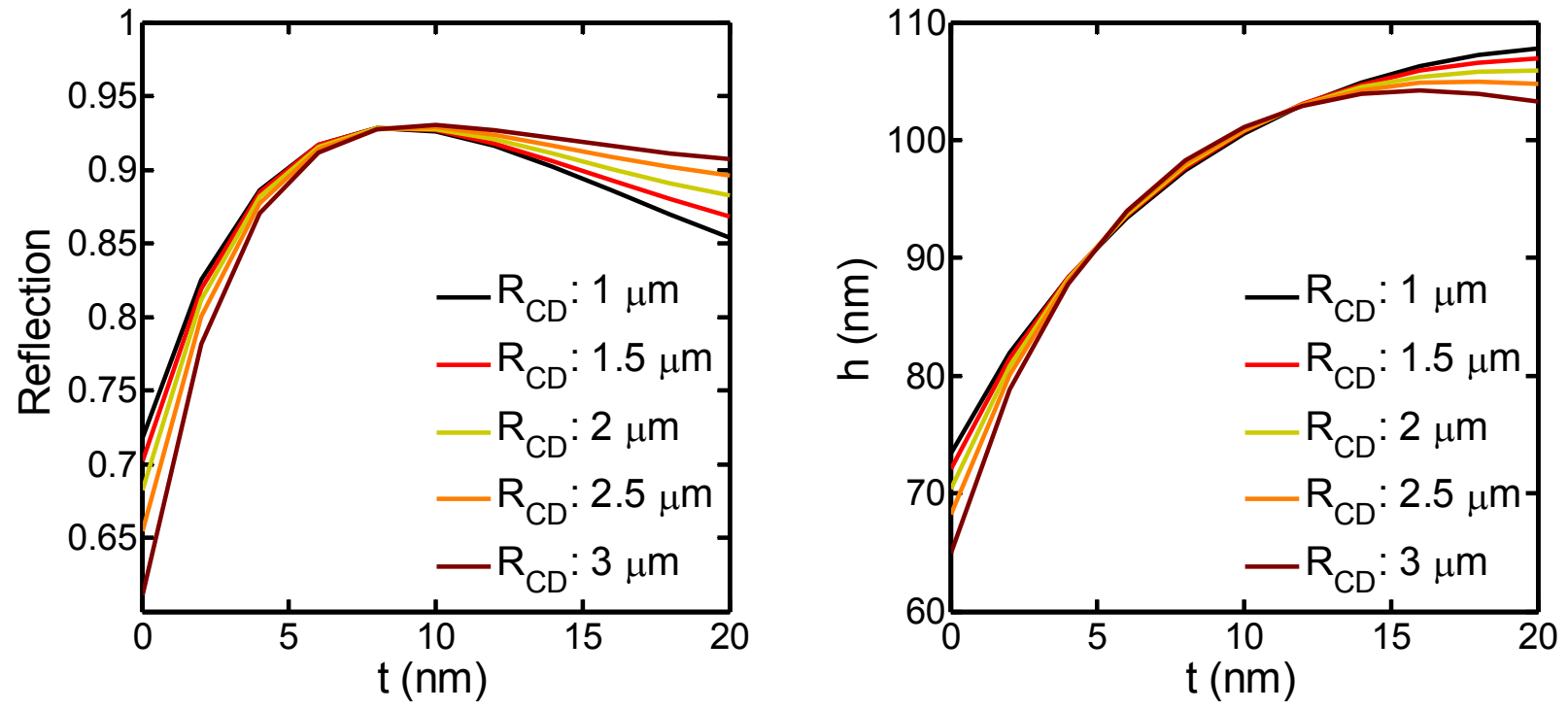

**Fig. 5. Reflection and *h* as function of *t* and the radius $R_{CD}$ of the computational domain. Parameters: Modes: 400 and $O_{max}$: 6**

The influence of the radius $R_{CD}$ of the computational domain is studied in Fig. 5. Whereas good convergence was achieved with respect to the number of modes and angular orders included in the calculation, the radius of the computational domain has a non-negligible influence on both the reflection and *h*.

The standard procedure for reducing the influence of the size of the computational domain is the implementation of an absorbing boundary condition, usually referred to as a perfectly matched layer (PML). The complex coordinate stretching formalism was implemented and the most simple version corresponding to the (f) curves in Figs. 3-4 of Ref. [6] was explored. The results are shown in Fig. 6.

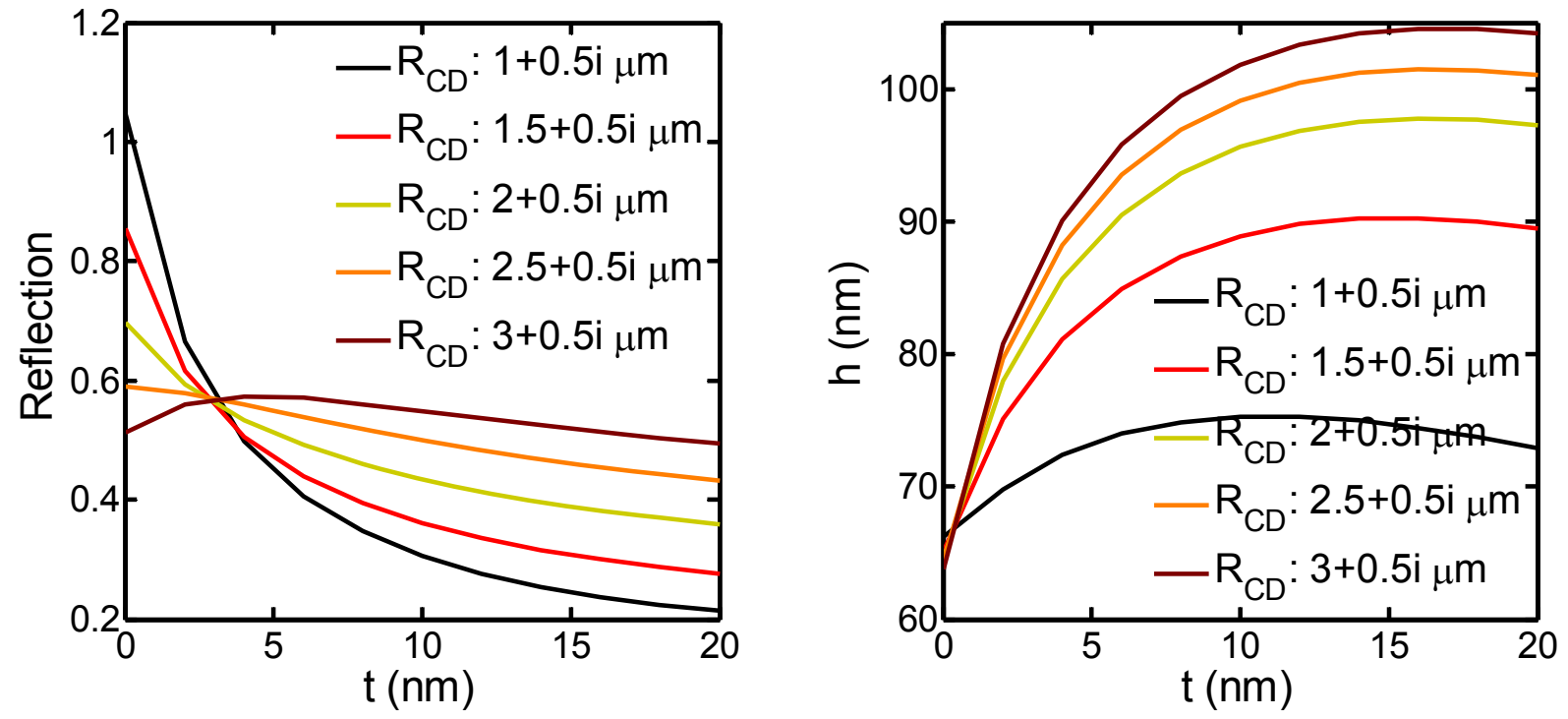

**Fig. 6. Reflection and *h* as function of *t* and the radius $R_{CD}$ of the computational domain. Parameters: Modes: 400 and $O_{max}$: 6**

The figure shows that the (f) implementation of PMLs does not reduce but rather amplifies the influence of the size of the computational domain. This influence was not studied in Ref. **Error! Bookmark not defined.**, and the performance of various PML coordinate transformations for the modal method remains an unexplored issue in the literature.

## 4.2 Top annular gold ring contact section

The practical implementation of the top annular gold ring contact may involve the implementation of a polymer surrounding the nanowire and additional planarization and lithography steps and may indeed prove quite challenging. However, from the optical engineering side the only design concern for the top gold contact is scattering of light due to the large index of the metal and the resulting reduction in efficiency.

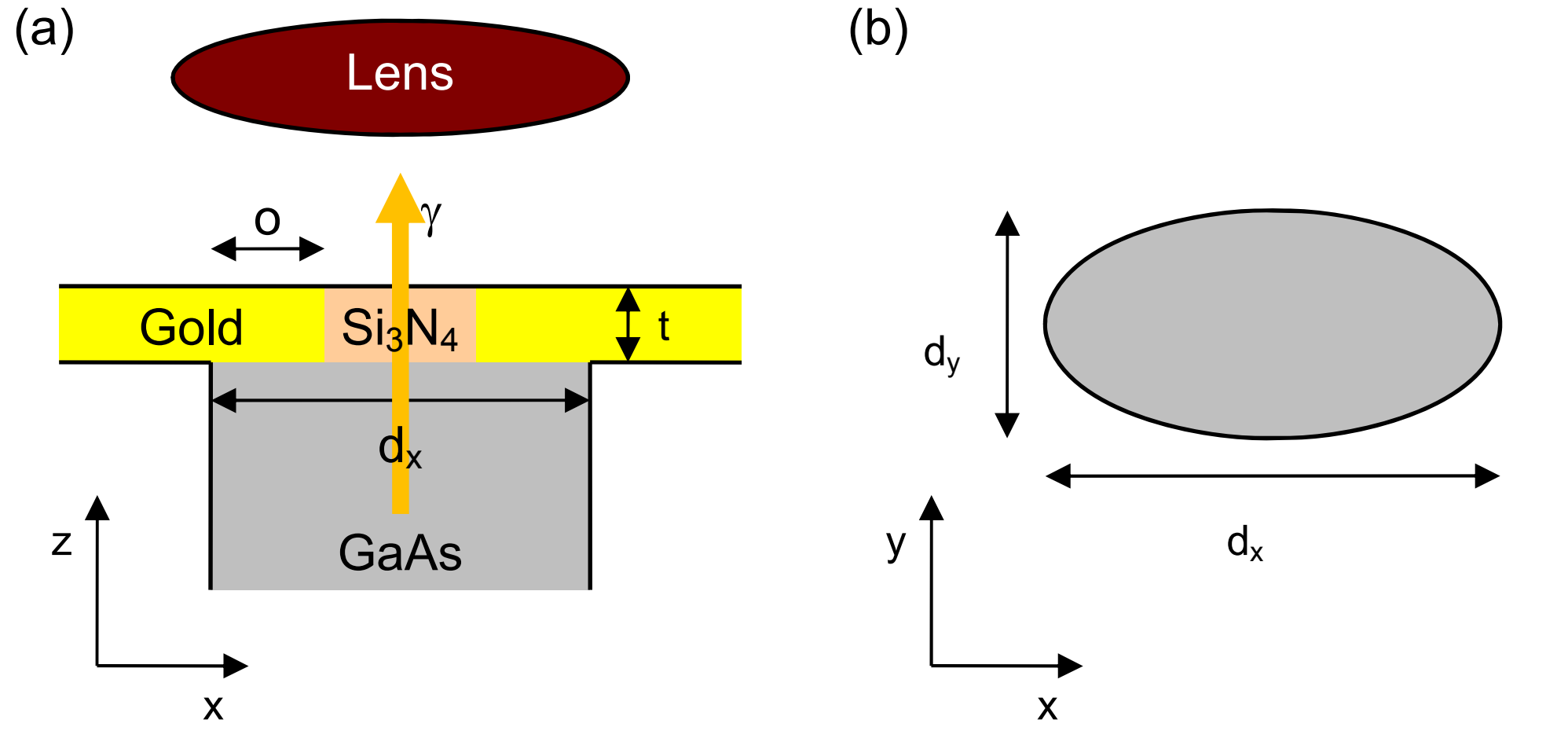

**Fig. 7. (a) Top part of the elliptical nanowire geometry. (b) Cross-section of the GaAs nanowire.**

Thus, the geometry under consideration is the very top part of the electrically contacted nanowire geometry illustrated in Fig. 7. It features the usual silicon nitride anti-reflection coating of thickness $t$ = $\lambda/(4n_{Si3N4})$ employed to reduce unwanted reflection at the semiconductor-air interface. Additionally, it features the annular gold ring which overlaps with the GaAs nanowire section with an overlap distance o which should be chosen sufficiently large to ensure good electrical conduction.

At the position of the quantum dot layer, the ideal geometrical parameters are $d_x$ = 280 and $d_y$ = 130 nm. Assuming that top-down etch defining the nanowire taper is isotropic, we have the relation

$$d_x = d_y + 150 \text{ nm}. \tag{120}$$

along the entire taper. For larger top diameters, the relative position of the focal point will move towards the origin, and the elliptical cross-section will appear more and more circular.

In the following, we study the total transmission $\gamma$ to a lens with a 0.8 numerical aperture as function of the top taper semi-diameter $d_y$, with $d_x$ given by Eq. (120). Unless otherwise stated, the convergence checks will be performed for a metal contact-nanowire overlap $o$ of 100 nm, for 200 modes per order, an angular order $O_{max}$ = 1 and a semidiameter of the computational domain $R_{CD}$ = 3 $\mu$m along the short $y$ semi-axis.

To verify the calculations, convergence checks are performed as function of the number of included modes per order and included angular orders $O_{max}$ in the calculation. Furthermore, a convergence check with respect to the radius $R_{CD}$ along the short $y$ semi-axis of the computational domain is performed.

The total transmission as function of $d_y$ for varying number of included modes is illustrated in Fig. 8(left). The curves overall agree reasonably well, however the convergence is not perfect for this range of included modes. From previous experience with the rotationally symmetric geometry analysed using the semi-analytical model, only around 150-200 eigenmodes are required for good convergence. The slow convergence here is not due to a lack of eigenmodes in itself but rather due to the general difficulty of resolving a large discontinuity in a field profile using a plane-wave type expansion.

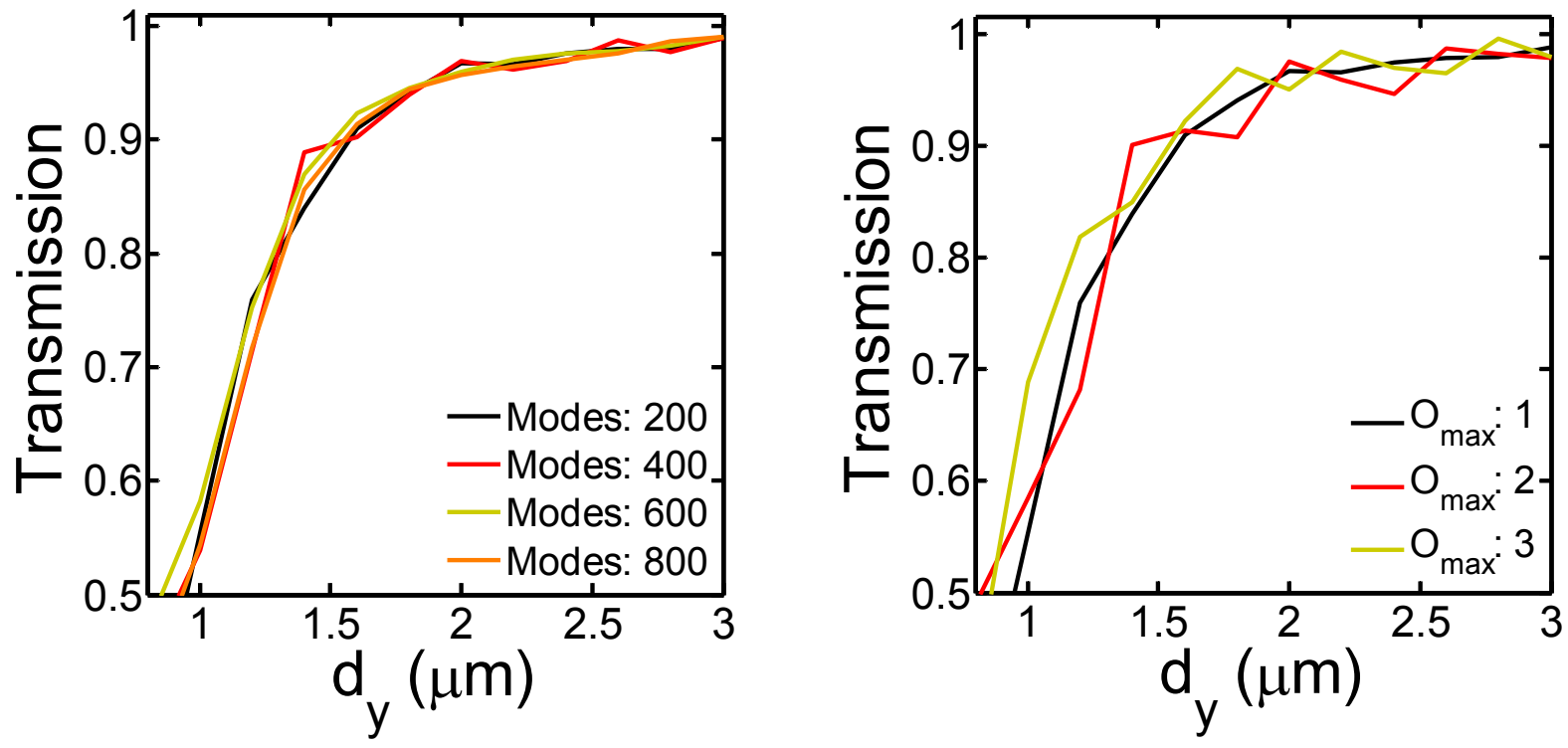

**Fig. 8. Transmission γ to a 0.8 NA lens as function of semi-diameter $d_y$ for varying number of modes (left) and for varying number of orders $O_{max}$ (right).**

The second relevant convergence check is with respect to the number of included orders $O_{max}$. Results for $O_{max}$ = 1, 2 and 3 are presented in Fig. 8(right). The curves for each order are somewhat irregular and display apparent oscillatory variations. It is not expected that these oscillations are physical but rather that they appear due to lack of convergence due to the discontinuity difficulty described above. Within some margin of error, we can say that the curves lie on top of each other. This is also confirmed by initial test calculations using the semi-analytical model in a geometry without metal, which could be performed correctly. The fact that only one order is required to correctly describe the transmission indicates that the elliptical shape in this parameter regime can be considered a weak perturbation to the perfectly rotationally symmetric structure.

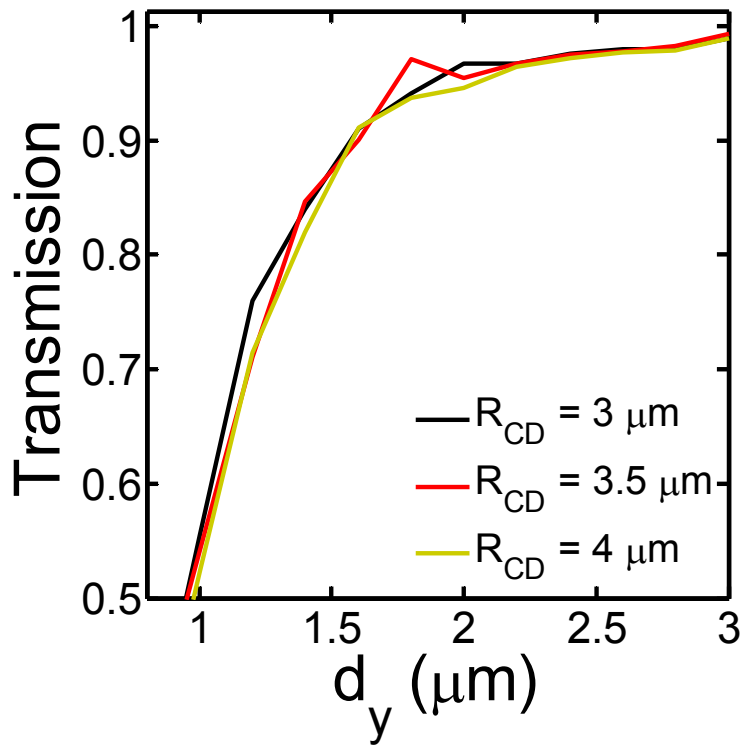

**Fig. 9. Transmission γ to a 0.8 NA lens as function of semi-diameter $d_y$ for varying radius $R_{CD}$ along the short semi-axis of the computational domain.**

A final convergence check is performed with respect to the size of the computational domain. In Fig. 9, the transmission is presented for three different radii $R_{CD}$ of 3, 3.5 and 4 μm. Again, the curves are not completely regular and again these irregularities are attributed not to size effects of the computational domain but rather from the discontinuity issue. For the fairly large top semi-diameters considered here, the light is preferentially emitted in the forward direction and artefacts from the limited lateral domain size are almost absent. This was again confirmed using initial test calculations with the semi-analytical model.

Even though perfect convergence is not obtained, the overall physical influence of the presence of the top annular gold ring contact can still be analysed.

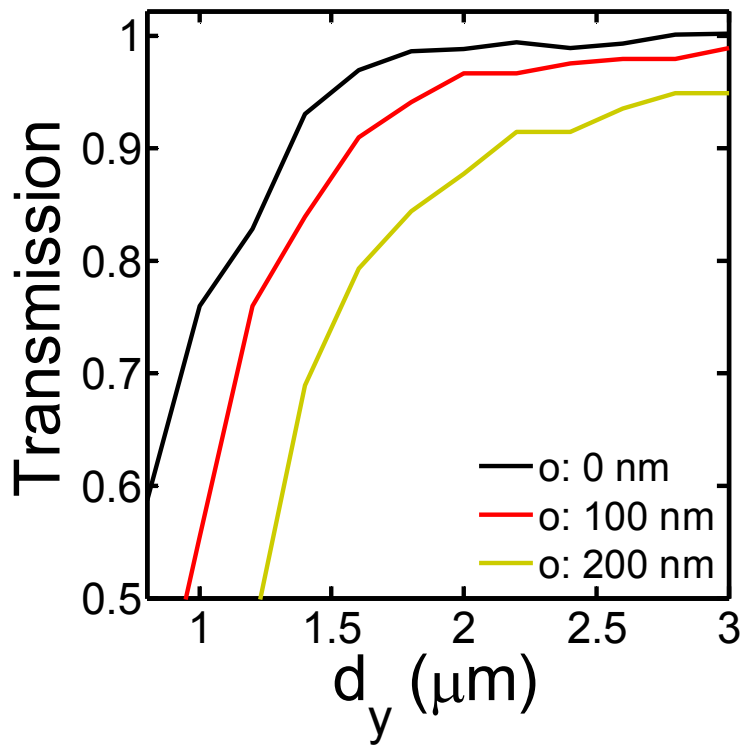

**Fig. 10. Transmission γ to a 0.8 NA lens as function of semi-diameter $d_y$ for varying metal-contact overlaps *o*.**

The transmission to the lens is presented in Fig. 10 for varying overlaps *o* of the gold contact with the GaAs elliptical nanowire. In the absence of overlap, the transmission increases for increasing semi-diameter $d_y$, and transmissions above 0.95 are obtained for a $d_y$ of 1.5 μm.

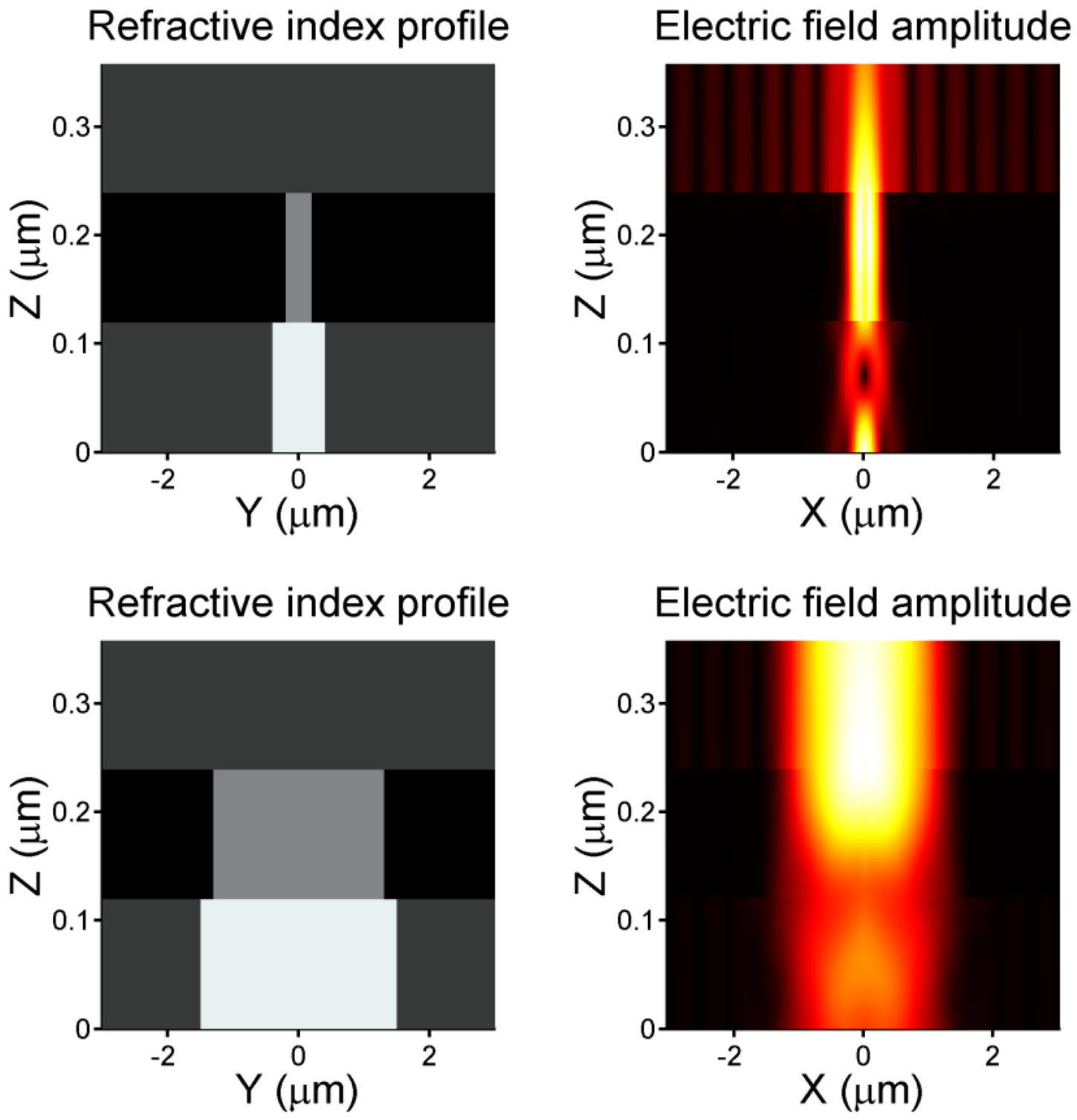

**Fig. 11. Refractive index profile (left) of the top contact section shown in Fig. 2(a) for $d_y$ = 0.8 μm (top) and $d_y$ = 3 μm (bottom) for *o* = 200 nm. Corresponding electrical field amplitude profiles (right).**

We now study the influence of a significant overlap of *o* = 200 nm. The refractive index and the electric field profiles for the top contact section presented in Fig. 7(a) are illustrated in Fig. 11 for a small and a large value of the semi-diameter $d_y$. For the small value $d_y$ = 0.8 μm, the overlap is relatively large and the fundamental mode is strongly scattered by the contact. However, for the large value $d_y$ = 3 μm, the relative overlap is decreased and we observe that the fundamental mode largely propagates unhindered through the contact section.

For a specific value of the overlap *o*, the relative overlap with the gold contact can thus be reduced by increasing the semi-diameter $d_y$, and a transmission of 0.95 can indeed be obtained by increasing the top semi-diameter $d_y$ to ~ 3 μm. This of course requires a longer taper height to achieve an adiabatic transition of the fundamental mode. However, an overlap *o* of only 100 nm or less is expected to be sufficient to achieve good electrical contacting, and in this case the transmission suffers less from the presence of the gold contact.

## 5 Acknowledgements

This work was funded by project SIQUTE (contract EXL02) of the European Metrology Research Programme (EMRP). The EMRP is jointly funded by the EMRP participating countries within EURAMET and the European Union.

## 6 References

1. J. Claudon, N. Gregersen, P. Lalanne, and J.-M. Gérard, "Harnessing light with photonic nanowires: Fundamentals and applications to quantum optics," ChemPhysChem **14**, 2393–2402 (2013).

2. M. Munsch, J. Claudon, J. Bleuse, N. S. Malik, E. Dupuy, J.-M. Gérard, Y. Chen, N. Gregersen, and J. Mørk, "Linearly polarized, single-mode spontaneous emission in a photonic nanowire," Phys. Rev. Lett. **108**, 077405 (2012).

3. A. V. Lavrinenko, J. Lægsgaard, N. Gregersen, F. Schmidt, and T. Søndergaard, *Numerical Methods in Photonics* (CRC Press, 2014).

4. L. Li, "Use of Fourier series in the analysis of discontinuous periodic structures," J. Opt. Soc. Am. A **13**, 1870–1876 (1996).

5. T. Häyrynen, J. R. de Lasson, and N. Gregersen, "Open geometry Fourier modal method: Modeling nanophotonic structures in infinite domains," J. Opt. Soc. Am. A, accepted.

6. J. P. Hugonin and P. Lalanne, "Perfectly matched layers as nonlinear coordinate transforms: a generalized formalization," J. Opt. Soc. Am. A **22**, 1844–1849 (2005).

7. N. W. McLachlan, *Theory and Application of Mathieu Functions* (Oxford University Press, 1964).

8. N. Gregersen, T. R. Nielsen, J. Mørk, J. Claudon, and J.-M. Gérard, "Designs for high-efficiency electrically pumped photonic nanowire single-photon sources," Opt. Express **18**, 21204–21218 (2010).

9. J. C. Gutiérrez-Vega, R. M. Rodríguez-Dagnino, M. A. Meneses-Nava, and S. Chávez-Cerda, "Mathieu functions, a visual approach," Am. J. Phys. **71**, 233–242 (2003).

10. I. Friedler, P. Lalanne, J. P. Hugonin, J. Claudon, J. M. Gérard, A. Beveratos, and I. Robert-Philip, "Efficient photonic mirrors for semiconductor nanowires," Opt. Lett. **33**, 2635–2637 (2008).